\documentclass[fleqn,usenatbib]{mnras}

\usepackage{newtxtext,newtxmath}

\usepackage[T1]{fontenc}

\DeclareRobustCommand{\VAN}[3]{#2}
\let\VANthebibliography\thebibliography
\def\thebibliography{\DeclareRobustCommand{\VAN}[3]{##3}\VANthebibliography}

\usepackage{graphicx}	
\usepackage{amsmath}	

\title{Milliarcsecond Core Size Dependence of the Radio Variability of Blazars}

\author[Hsu et al. 2023]{
Po Chih Hsu,$^{1,2}$ Jun Yi Koay,$^{2}$ Satoki Matsushita,$^{2}$ Chorng-Yuan Hwang,$^1$ Talvikki Hovatta,$^{3,4}$ \newauthor
Sebastian Kiehlmann,$^{6,7}$ Anthony Readhead,$^{5}$ Walter Max-Moerbeck$^8$ and Rodrigo Reeves$^{9}$ 
\\
$^{1}$Graduate Institute of Astronomy, National Central University, 300 Zhongda Rd., Zhongli, Taoyuan 32001, Taiwan (R.O.C.)\\
$^{2}$Institute of Astronomy and Astrophysics, Academia Sinica, Taipei 10617, Taiwan (R.O.C.)\\
$^{3}$Finnish Centre for Astronomy with ESO (FINCA), University of Turku, FI-20014 Turku, Finland \\
$^{4}$Aalto University Mets\"{a}hovi Radio Observatory, Mets\"{a}hovintie 114, FI-02540 Kylm\"{a}l\"{a}, Finland \\
$^{5}$Owens Valley Radio Observatory, California Institute of Technology, Pasadena, CA 91125, USA \\
$^{6}$Institute of Astrophysics, Foundation for Research and Technology-Hellas, GR-71110 Heraklion, Greece \\
$^{7}$Department of Physics, University of Crete, GR-70013 Heraklion, Greece \\
$^{8}$Departamento de Astronom\'{i}a, Universidad de Chile, Camino El Observatorio 1515, Las Condes, Santiago, Chile \\
$^{9}$ Departamento de Astronom\'{i}a, Universidad de Concepci\'{o}n, Concepci\'{o}n, Chile \\
}

\date{}

\pubyear{2023}

\begin{document}
\label{firstpage}
\pagerange{\pageref{firstpage}--\pageref{lastpage}}
\maketitle

\begin{abstract}
    Studying the long-term radio variability (timescales of months to years) of blazars enables us to gain a better understanding of the physical structure of these objects on sub-parsec scales, and the physics of super massive black holes. 
    In this study, we focus on the radio variability of 1157 blazars observed at 15~GHz through the Owens Valley Radio Observatory (OVRO) Blazar Monitoring Program.
    We investigate the dependence of the variability amplitudes and timescales, characterized based on model fitting to the structure functions, on the milliarcsecond core sizes measured by Very Long Baseline Interferometry. 
    We find that the most compact sources at milliarcsecond scales exhibit larger variability amplitudes and shorter variability timescales than more extended sources. 
    Additionally, for sources with measured redshifts and Doppler boosting factors, the correlation between linear core sizes against variability amplitudes and intrinsic timescales are also significant. 
    The observed relationship between variability timescales and core sizes is expected, based on light travel-time arguments.
    This variability vs core size relation extends beyond the core sizes measured at 15\,GHz; we see significant correlation between the 15\,GHz variability amplitudes (as well as timescales) and core sizes measured at other frequencies, which can be attributed to a frequency-source size relationship arising from the intrinsic jet structure. At low frequencies of 1\,GHz where the core sizes are dominated by interstellar scattering, we find that the 
    variability amplitudes have significant correlation with the 1~GHz intrinsic core angular sizes, once the scatter broadening effects are deconvoluted from the intrinsic core sizes.
\end{abstract}


\maketitle

\section{Introduction}
\label{introduction}

Active Galactic Nuclei (AGNs) are compact regions at the center of galaxies, powered by accretion onto supermassive black holes (SMBHs), emitting radiation across the electromagnetic (EM) spectrum from radio to gamma-rays. 
Due to their very compact nature, AGNs are variable objects at all wavelengths, e.g., in radio \citep{2004A&A...419..485C, 2007A&A...462..547F, 2007A&A...469..899H, 2009AJ....137.5022N}, in optical \citep{1998ApJ...504..671K, 2016ApJ...827...53W, 2018MNRAS.476.2501A}, and in X-rays \citep{2009A&A...504...61I, 2012ApJ...746...54G, 2016ApJ...831..145Y}.

At radio wavelengths, the emission of the AGN is typically dominated by non-thermal synchrotron radiation from the relativistic jets, particularly in radio-loud AGNs. The intrinsic radio variability of AGNs can originate from various processes, such as shocks produced in the jet \citep{1985ApJ...298..301H, 2008A&A...485...51H}, jet precession that can lead to periodic variations \citep{2011A&A...526A..51K}, variable accretion rates due to disk instability \citep{1986ApJ...305...28L}, tidal disruption events caused by sudden accretion of the star debris or a star rotating into the tidal sphere of a SMBH
\citep{2015aska.confE..54D}, and last for few months \citep{2020SSRv..216...81A}, as well as changes of jet Doppler boosting factors \citep{1992A&A...259..109G, 1992A&A...255...59C}. Studying the intrinsic radio variability of AGNs thereby enable us to better understand the physical structures of the radio jets and allow us to probe the innermost regions and the physics of SMBH in the center of galaxies.

For the most compact AGNs, observed radio variability can also arise due to extrinsic propagation effects, as the radio waves emitted by the AGN travel through various media towards the observer.
Centimeter wavelength intra-day variability (IDV) in compact AGNs was first observed by \citet{1984AJ.....89.1111H}.
It is now recognized that interstellar scintillation (ISS, \citealp{ 1987AJ.....93..589H, 1990ARA&A..28..561R, 2008ApJ...689..108L, 2018MNRAS.474.4396K}) is mainly responsible for IDV of blazars at cm-wavelengths. 
The turbulent, ionized ISM of our Galaxy scatters radio waves propagating through it. 
Transverse motions between the scattering material and the Earth causes rapid flux density fluctuations to be observed as regions of constructive and destructive interference drift across the observing telescope's field of view.
The evidence for ISS comes from the observed annual cycle of IDV timescales \citep{2001ApJ...550L..11R, 2001A&A...370L...9J, 2003A&A...404..113D, 2003ApJ...585..653B}, due to the Earth's orbital motion about the Sun causing periodic relative motion to the ISM.    
The variability of blazars is known to be dominated by ISS at cm-wavelengths \citep{2008ApJ...689..108L, 2018MNRAS.474.4396K}. 

Blazars, which by their very nature have relativistic jets pointed almost directly towards the Earth, are ideal candidates for studying the intrinsic radio variability of AGNs, particularly in relation to jet properties.
Due to Doppler beaming \citep{1978PhyS...17..265B}, their variability timescales are compressed, exhibiting flux density variations on timescales of weeks to years, well within human lifetimes. 
The compact nature of blazars also predisposes them to exhibiting ISS, which can be the dominant cause of variability at lower frequencies of $\lesssim$ 15\,GHz and at short timescales of days to weeks \citep{2011AJ....142..108K, 2019MNRAS.489.5365K}; the effect of ISS is also dependent on Galactic latitudes \citep{2008ApJ...689..108L}.

Blazars can be divided into two sub-classes: BL Lac objects (BLOs), and flat spectrum radio quasars (FSRQs). 
FSRQs typically exhibit broad optical emission lines, which are attributed to the thermal disk and broad-line region dominating their optical spectra \citep{2011A&A...525L...8C}.
In contrast, BLOs are dominated by continuum synchrotron radiation in the optical regime \citep{2002ApJ...564...92C} and either lack such lines or have very weak ones.
Additionally, FSRQs are generally more luminous than BLOs.

There have been many studies focusing on AGN variability using small samples of sources with good time sampling rates (e.g., \citealp{2013A&A...552A..11R, 2021A&A...654A..38P}).
There are also large sample AGN variability studies, although with very sparse time sampling.
These studies used several multi-epoch radio survey datasets, including that of the Murchison Widefield Array \citep{2014MNRAS.438..352B, 2020bugm.conf...57R}, the Australia Telescope Compact Array \citep{2015MNRAS.450.4221B, 2016MNRAS.461.3314H}, the Karl G. Jansky Very Large Array \citep{2011ApJ...737...45O, 2013ApJ...769..125H}, and the Allen Telescope Array \citep{2011ApJ...739...76B}, to search for variable sources and slow transients.
However, there is a lack of large-sample studies of blazar variability with good sampling rate of AGN variability. \
The Monitoring of Jets in Active Galactic Nuclei with VLBA Experiments (MOJAVE, \citealt{2005AJ....130.1389L}) survey is the first long-term observation of large sample of AGNs, aiming to study the structure and evolution of relativistic outflows.

The Owens Valley Radio Observatory (OVRO) blazar monitoring program \citep{2011ApJS..194...29R} has obtained the lightcurves of more than 1500 blazars over the past 15 years (beginning in late 2007), each about twice per week, enabling studies of the radio variability of a large sample blazars with very dense time sampling. 
From the OVRO data, \citet{2011ApJS..194...29R} found that gamma-ray loud blazars have almost a factor of two higher variability amplitudes than gamma-ray quiet sources, at a high significance level ($>6\sigma$). 
This result was also confirmed in the follow-up paper by \citet{2014MNRAS.438.3058R}.
They also found that BLOs typically have larger variability amplitudes than FSRQs.
Thirdly, low redshift FSRQs ($z < 1$) are significantly more variable than high z FSRQs, attributed to time dilation effects.
ISS has also recently been found to be present in the OVRO lightcurves on the shortest 4-day timescales \citep{2019MNRAS.489.5365K}. 

One intriguing aspect that has yet to be examined using the large OVRO dataset is the relationship between intrinsic blazar radio variability and the compactness of the source. 
\citet{2005AJ....130.2473K} studied 250 flat-spectrum extragalactic radio sources observed with VLBI at sub-milliarcsecond scales and found that sources exhibiting IDV tend to have more compact core sizes and are more core-dominated than sources that do not exhibit IDV.
\citet{2004ApJ...614..607O} similarly showed that the core-dominated blazars at milliarcsecond scales are more likely to show scintillation-induced IDV at cm-wavelengths in comparison to less compact or core-dominated sources.
These studies were mainly focused on the relationship between IDV, and core compactness. In a more recent study, \citet{2018MNRAS.474.4396K} found a significant correlation between the 15\,GHz variability modulation index (representing long-term intrinsic variability amplitudes) and the milliarcsecond angular core sizes of a small subset of about 100 OVRO-monitored blazars.

In this study, we make use of the OVRO data to further examine the dependence of intrinsic blazar variability amplitudes and timescales on the milli-arcsecond core sizes as measured using Very Long Baseline Interfometry (VLBI) observations.  
This will be for a larger sample of source core sizes than studied previously, extending the work of \citet{2018MNRAS.474.4396K} with more detailed analysis. 
This study also focuses mainly on the long-term intrinsic variability, as opposed to IDV as in previous studies, which are likely to be dominated by ISS, even at 15\,GHz \citep{2019MNRAS.489.5365K}.
Section~\ref{Data} describes the observational data used in this study. 
In Section~\ref{Data Analysis and Results}, we describe how we characterize the source variability amplitudes and timescales based on the structure function (SF), as well as present the main results examining their relationship with the milliarcsecond core sizes measured at multiple wavelengths.
We also study the effects of interstellar scattering on the variability vs. core size relationship at low frequencies of 1\,GHz, where scattering effects dominate.
Lastly, section~\ref{conclusion} summarizes our main results and discusses future plans.

\section{Data}    
\label{Data}

There are two main data sets that we used for this work, which we describe in this section.
We used data from the 15 GHz OVRO Blazar Monitoring Program \citep{2011ApJS..194...29R} to characterize blazar variability amplitudes and timescales.
To study blazar variability dependence with core sizes, we make use of VLBI core sizes measured at 8 different frequencies (i.e., 1, 2, 5, 8, 15, 24, 43, and 86~GHz) derived by \citet{2022MNRAS.515.1736K}.

\subsection{The 15 GHz OVRO Blazar Monitoring Program} 
\label{OVRO}

Light curve data were obtained from the 15\,GHz OVRO Blazar Monitoring Program \citep{2011ApJS..194...29R}, where the flux densities of about 1500 blazars have been measured twice per week since 2008 to the present, using the OVRO 40-m single dish telescope, with minimum sensitivity of 4 mJy, and with typical 3\% uncertainties. 
The median sampling cadence of a source is about 4 days.
The initial 1157 sources monitored by OVRO were selected from the Candidate Gamma-Ray Blazar Survey (CGRaBS, \citealp{2008ApJS..175...97H}) that are North of $-20 \degr$ declination.
CGRaBS is statistically selected from the Combined Radio All-Sky Targeted Eight GHz Survey (CRATES, \citealp{2007ApJS..171...61H}) and can therefore be considered as a complete sample.
Sources with radio spectral indices larger than $-$0.5 were selected, based on 1~GHz data from the NRAO VLA Sky Survey (NVSS, \citealp{1998AJ....115.1693C}) and 8.4~GHz data from the Sydney University Molonglo Sky Survey (SUMSS, \citealp{1999AJ....117.1578B}).
In this paper, we are interested in the variability characteristics of radio-selected of AGNs; we thus ignore the additional 400 gamma-ray selected sources in the OVRO sample.

\subsection{VLBA Core Size Measurements at Eight Frequencies}
\label{VLBA Core Size Measurements at Eight Frequencies} 

VLBI observations of blazars typically reveal a core-jet structure, where the extended jet components are typically optically thin, while the unresolved, compact core is optically thick.
At centimeter wavelengths, the blazar core is interpreted as the $\tau \sim 1$ surface of the jet.
As we move to higher frequencies, the observations are then probing regions further upstream of the jet, and at mm-wavelengths, may probe regions close to the jet-launching region. 
Therefore, the term `core size' in VLBI observations typically refer to the size of this unresolved, compact core region, associated with the opaque region of the jet at cm-wavelengths.
For our study, we want to examine the relationship between the intrinsic variability of blazars and the measured sizes of these compact cores, assuming that the observed flux density variations in the OVRO lightcurves do indeed physically correspond to changes within the core component, particularly at 15\,GHz.

We used blazar core size measurements compiled and derived in the paper by \citet{2022MNRAS.515.1736K}, which extends the work of an earlier paper by \citet{2015MNRAS.452.4274P}. 
 \citet{2022MNRAS.515.1736K} provide the largest sample of sources size measurements at milliarcsecond VLBI scales, increasing the sample of core size measurements by a factor of 3 times that of the precursor paper by \citet{2015MNRAS.452.4274P}.
Sources are initially selected from the Astrogeo data base, including more than 17,000 sources measured at multiple radio frequencies.
\citet{2022MNRAS.515.1736K} estimated the VLBI-scale core sizes of 157, 4878, 13235, 15991, 751, 501, 257, and 113 sources at 1, 2, 5, 8, 15, 24, 43, and 86 GHz respectively.
Core size determinations with uncertainties greater than 50 \% were discarded by the authors, thus giving a remaining sample of 8959 sources that passed the criteria and have core sizes determined at at least one frequency. 
The lower frequencies, i.e., 2, 5, and 8~GHz (excluding 1~GHz) contain the largest samples of core size measurements.
Their core sizes were estimated using multi-epoch archival survey data from 1994 to 2021, observed using the Very Long Baseline Array (VLBA). 
To do this, the authors fitted two circular Gaussian component models to the self-calibrated VLBA visibilities of each epoch for their sources \citep{2005AJ....130.2473K, 2022MNRAS.515.1736K}, one for the core, and the other for the extended jet emission, assuming a one-sided core-jet structure for these blazars. 
For our analyses in this paper, we use the median value of the core size of each source as a representative core size.

Redshift measurements were compiled from both \citet{2014MNRAS.438.3058R} and \citet{2015MNRAS.452.4274P}.
Table~\ref{OVRO core size sample number} summarizes the total number of sources, $N_\mathrm{\theta}$, that have core sizes measurements at each frequency in the paper by \citet{2022MNRAS.515.1736K}, the sub-sample of sources that have core size measurements at each frequency that also have OVRO lightcurves ($N_\mathrm{\theta, \ OVRO}$), and the sample of sources that also have redshift measurements $N_\mathrm{\theta, \ OVRO, \ z}$. We use the latter two samples in our analysis.
Note that the numbers presented here include the 16 non-variable sources, which we later remove from our analysis (see section~\ref{sssec: Structure Function Fitting}).

\begin{table}
    \begin{center}
        \caption{Multi-frequency core sizes sample from \citet{2022MNRAS.515.1736K} that overlapped with OVRO \citet{2011ApJS..194...29R} sample in the analysis.
        }
        \begin{tabular}{|c|c|c|c|} \hline
            Frequency (Band) & $N_\mathrm{\theta}$ & $N_\mathrm{\theta, \ OVRO}$ & $N_\mathrm{\theta, \ OVRO, \ z}$ \\  
            (1) & (2) & (3) & (4) \\  \hline\hline
            1~GHz (Band~L) & 135 & 116 & 109 \\
            2~GHz (Band~S) & 3541 & 865 & 768 \\
            5~GHz (Band~C) & 4543 & 464 & 424 \\
            8~GHz (Band~X) & 7039 & 1053 & 917 \\
            15~GHz (Band~U) & 670 & 376 & 346 \\
            24~GHz (Band~K) & 259 & 156 & 147 \\
            43~GHz (Band~Q) & 205 & 158 & 150 \\
            86~GHz (Band~W) & 48 & 1 & 1 \\  \hline
        \end{tabular}
        {\parbox{3 in}{
        Note: (1) core size sample number \citep{2022MNRAS.515.1736K};  (2) core size sample overlapped with OVRO \citep{2011ApJS..194...29R}; (3) core size sample overlapped with OVRO and with redshift measurement \citep{2014MNRAS.438.3058R, 2015MNRAS.452.4274P} }
        }
        \label{OVRO core size sample number}
    \end{center}
\end{table}

In sections~\ref{ssec: Dependence of Variability Amplitudes on Observed/Linear Core Sizes} and \ref{Dependence of Variability Timescale on Observed/Linear Core Sizes}, we mainly focus on the observed core sizes at frequencies between 2 to 43\,GHz from \citet{2022MNRAS.515.1736K}. 
Core sizes observed at 86 GHz were excluded, since there is only a single source that overlaps with the OVRO sample.
1~GHz core sizes are significantly affected by scatter broadening by the ISM \citep{2022MNRAS.515.1736K}, and will be further analysed and discussed in section~\ref{Dependence of Variability Amplitudes and Timescales on Intrinsic/Scattering/Observed Core Sizes}.

\section{Data Analysis and Results} 
\label{Data Analysis and Results}
    
In this section, we first discuss how we characterized the blazar variability amplitudes and timescales (Section~\ref{ssec: Characterizing Variability Amplitudes and Timescales}).
We then show our analysis of the relationship between variability amplitudes with core sizes in section~\ref{ssec: Dependence of Variability Amplitudes on Observed/Linear Core Sizes}, and variability timescales with core sizes in section~\ref{Dependence of Variability Timescale on Observed/Linear Core Sizes}.
Next, we investigate how interstellar scattering affects the dependence of the variability amplitudes and timescales on 1~GHz core sizes in section~\ref{Dependence of Variability Amplitudes and Timescales on Intrinsic/Scattering/Observed Core Sizes}.

\subsection{Characterizing Variability Amplitudes and Timescales} 
\label{ssec: Characterizing Variability Amplitudes and Timescales}

\begin{figure}
    \centering
    \includegraphics[width=\columnwidth]{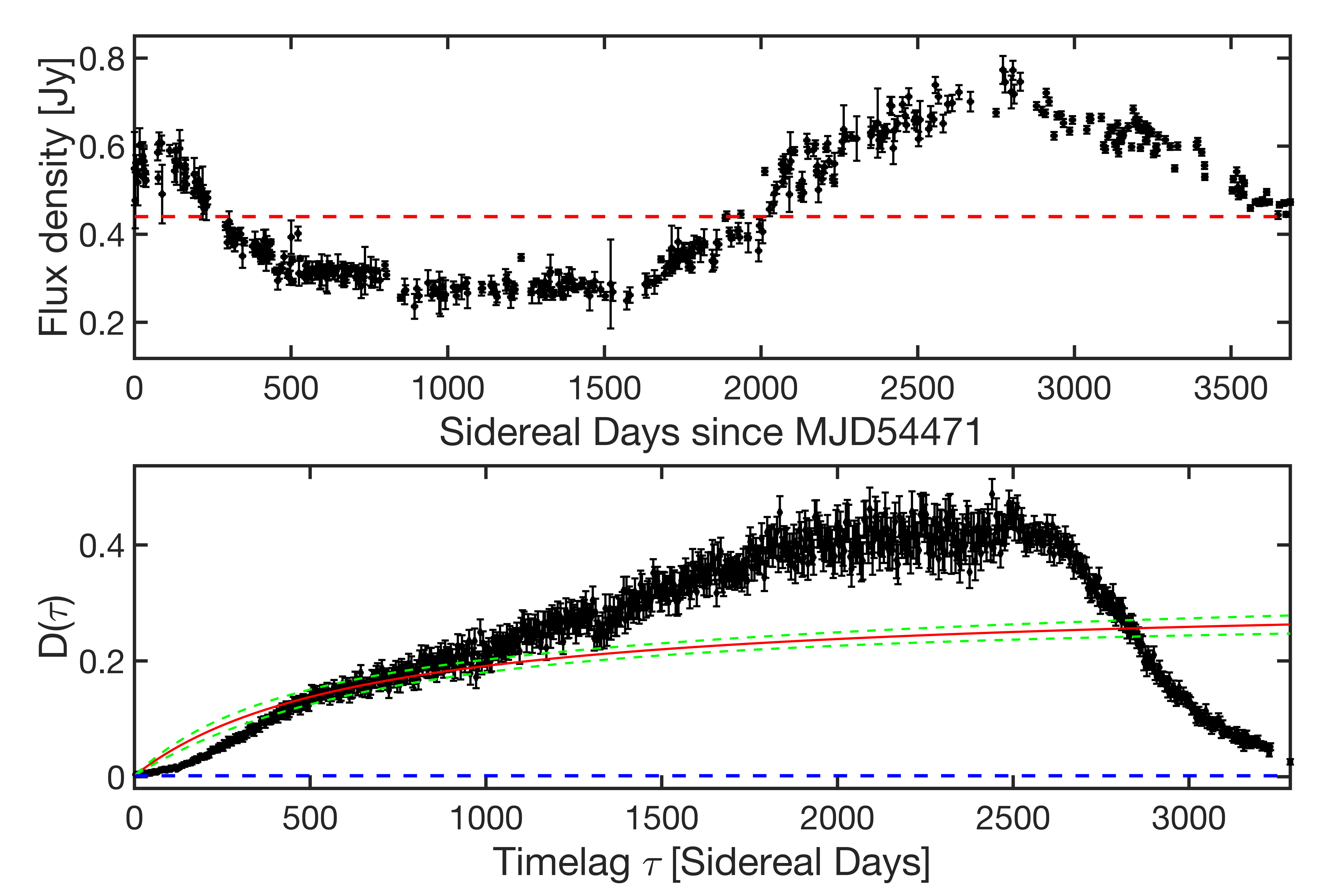}
    \includegraphics[width=\columnwidth]{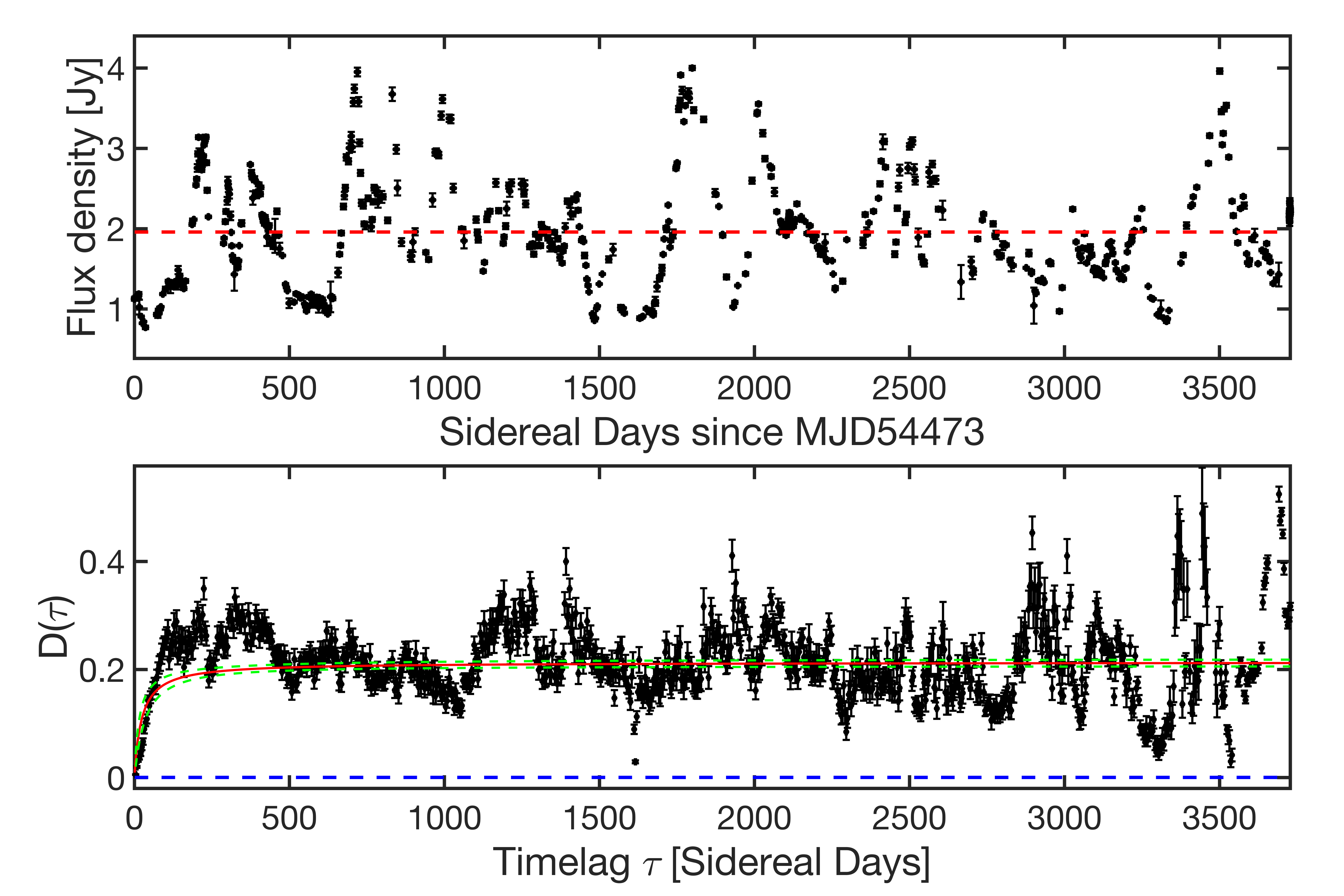}
    \caption{
    The 15\,GHz OVRO light curves and SFs of blazar sources J0502+1338 (upper panel) and J0721+7120 (lower panel); both sources have been observed over ten years, with the former displaying slow variations on the order of a few years, and the latter exhibiting faster timescale variations on the order of months. 
    The horizontal red dashed line in light curve panel shows the mean flux density of the sources.
    The red solid line in the SF panel shows the model fit to the SF, the pink dashed line shows 95 \% confidence bounds for the model fit, and the blue dashed line shows the $D_{\rm noise}$ (close to zero) contribution to the SF amplitude (see section~\ref{sssec: Structure Function Calculation} for details).
    }
    \label{lightcurves and SF}
\end{figure}

Fig~\ref{lightcurves and SF} shows the 15\,GHz lightcurves of two sources from the OVRO Blazar Monitoring Program \citep{2011ApJS..194...29R}. Some sources like J0721$+$7120 (lower panel of fig~\ref{lightcurves and SF}) exhibit variability on timescales on the order of weeks and months, while other sources such as J0502$+$1338 (upper panel of fig~\ref{lightcurves and SF}) show slower variations on the order of years. 
To characterize and quantify the variability characteristics of the OVRO sources, we first derive the SF from the lightcurves of each source, then perform a model fit to the SFs, from which we extract the variability timescale and variability amplitude of each source.

\subsubsection{Deriving the Structure Function (SF)} 
\label{sssec: Structure Function Calculation}
We used the SF to characterize the variability of the sources, following \citet{2008ApJ...689..108L}, \citet{2011AJ....142..108K} and \citet{2019MNRAS.489.5365K}. 
The SF is insensitive to the gaps in the sampling of data. 
The other advantage of using the SF is that it is not sensitive to biases resulting from errors in the estimation of the mean flux density of the sources.
The observed SF at a given time lag $\tau$ is given by:
\begin{equation}
    \label{structure function}
    D_{\rm obs}(\tau) = \frac{ 1 }{ N_{\tau} } \sum_{j,k}[S(t_{\rm j}) - S(t_{\rm k}-\tau)]^2
\end{equation}
where $D_{\rm obs}(\tau)$ is SF value, which is a dimensionless quantity, $S(t)$ is the measured flux density at time $t$, normalized by the mean flux density, $N_{\rm \tau}$ is the number of pairs of flux densities with a time lag $\tau$, binned to the nearest integer multiple of 4 days, which is the typical cadence between flux density measurements for each source.

The uncertainties ($D_{\rm err}$) of the SF amplitude at each time lag in figure~\ref{lightcurves and SF} and \ref{J1350+0940} were calculated using the following equation,
\begin{equation}
    \label{structure function error}
    D_{\rm err} = \frac{ \sigma_{\rm SF}} {\sqrt{N_{\rm \tau}-1}} \,
\end{equation}
where $\sigma_{\rm SF}$ is the standard deviation of $[S(t_{\rm j}) - S(t_{\rm k}-\tau)]^2$ in each time lag bin, and $N_{\rm \tau}$ is the pairs of the time lags.

\subsubsection{Structure Function (SF) Fitting} 
\label{sssec: Structure Function Fitting}
For a random stochastic process, when the observation duration is longer than its characteristic timescale, we can expect that $D_{\rm obs}(\tau)$ will increase with the time lag and saturate at a value two times that of the modulation index, $m$, of the source light curve.
Therefore, we can use a simple model to fit the SF \citep{2008ApJ...689..108L}, which is given by
\begin{equation}
    \label{structure function mod}
    D_{\rm mod}(\tau) = 2m^{2}  \left( \frac{ \tau }{\tau+\tau_{\rm char}} \right) + D_{\rm noise} ,
\end{equation}
where $2m^2$ is the value where the SF saturates.
$\tau_{\rm char}$ is the characteristic timescale, where the SF reaches half of its maximum value at saturation. 
$D_{\rm noise}$ is a constant additive noise floor due to systematic and instrumental errors that contribute to the overall variability, introducing a positive bias to the observed SF amplitudes.
Here we assume white noise that is constant across all time-lags.

We determine $D_{\rm noise}$ for each source based on the light curve data, given by:
\begin{equation}
    \label{Dnoise}
    D_{\rm noise} = 2 \left( \frac{ \tilde{S}_{\rm err} }{ \overline{S}} \right)^2 \,
\end{equation}
where $\tilde{S}_{\rm err}$ is the median flux density uncertainty for all data points in the light curve, and $\overline{S}$ is the mean flux density of the source.

For our analysis, we want to use the noise-debiased SF amplitude at 1000 days, $D(\rm 1000d)$, derived from the model fit, to characterize the source long term variability amplitudes. 
We selected $D(\rm 1000d)$ to ensure that the time lag is sufficiently large such that the SF already saturates in the majority of sources, and that the variability amplitudes are no longer dominated by ISS. At the same time, the time lag of 1000 days is not too large such that the SF amplitudes are dominated by measurement uncertainties arising from insufficient statistical sampling of the flux density variations at the longest timescales.
To enable us to directly extract $D(\rm 1000d)$, we use the following equivalent equation to perform the SF fitting:
\begin{equation}
    \label{SF 1000 days}
    D_{\rm mod}(\tau) = D(\rm 1000d)  \frac{1+\tau_{\rm char}/1000 }{1+\tau_{\rm char}/\tau}+ D_{\rm noise} \,.
\end{equation}
such that $D(\rm 1000d)$ become a fitted parameter instead of $2m^2$. This then gives us $D(\rm 1000d) = D_{\rm mod}(\rm 1000d)-D_{\rm noise}$, so that we obtain the noise-debiased SF amplitude.
When fitting the model, each $D_{\rm obs}(\tau)$ data point is weighted by the inverse of the error of the SF amplitude given by equation~\ref{structure function error}, the weight ($w$) is given by
\begin{equation}
    \label{weight}
    w = \frac{1}{D_{\rm err}(\tau) \left( \sqrt{ \frac{\tau}{\tau_{\rm max}}  } \right) \, }
\end{equation}
$\tau_{\rm max}$ is the entire observing span of the source. 
We multiply an additional $\sqrt{\tau /{\tau_{\rm max}}}$ term in the denominator, which places a stronger weight on the shorter time lags and to down-weighted the longer time lags. We found that this provided better model fits to the observed SF datapoints at timelags $\lesssim 1500$ days, including at 1000 days which we are focusing on.
We consider the uncertainties of $D(\rm 1000d)$ and $\tau_{\rm char}$ to be given by the 95\% confidence bounds ($\sim2 \sigma$) of the SF model fit.

\begin{figure}
    \centering
    \includegraphics[width=\columnwidth]{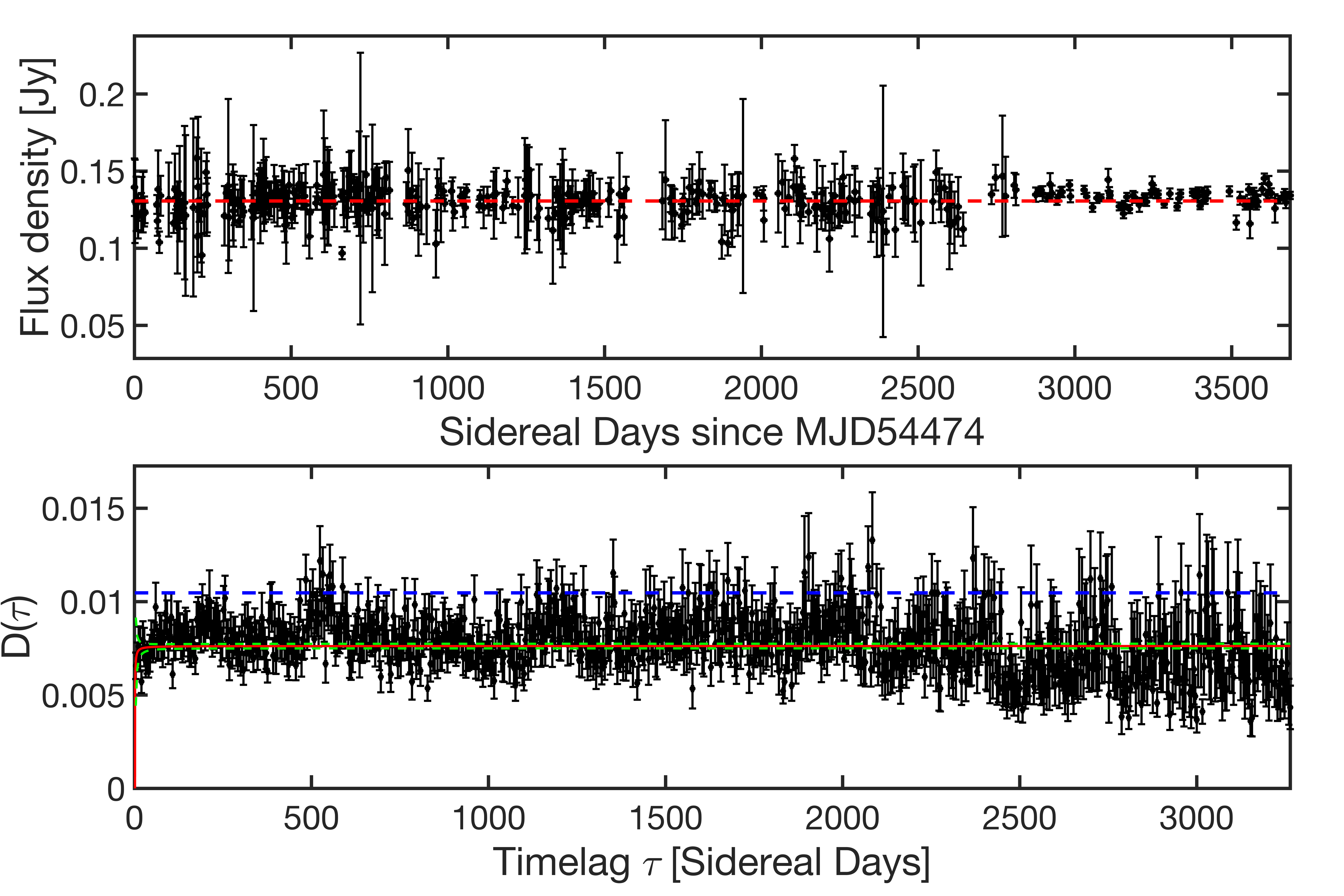}
    \caption{
    The 15\,GHz OVRO light curve and SF of the blazar source J1350+0940.
    This source does not appear to be variable, relative to the noise levels.
    Therefore, the model fitted $D_{\rm mod}(\tau)$ is typically lower than $D_{\rm noise}$ (blue dashed line) across all timescales. The SF model fitting therefore provides a negative estimate of $D(\rm 1000d)$ (red solid line) and the estimated $\tau_{\rm char}$ is not valid. We remove such sources from our analysis (see section~\ref{sssec: Structure Function Fitting} for details).
    }
    \label{J1350+0940}
\end{figure}

Fig~\ref{lightcurves and SF} shows two examples of the estimated SF and the corresponding SF model fits for the sources J0721$+$7120 (lower panel of fig~\ref{lightcurves and SF}) and J0502$+$1338 (upper panel of fig~\ref{lightcurves and SF}). 
In some sources that vary on timescales of months (e.g., J0721$+$7120), their SF amplitudes increase rapidly and saturate, and fluctuate over time after saturation, possibly due to some quasi-periodicity in the light curve. 
In 493 out of 1157 sources in our sample, the variability timescales are longer than the total observing time span, such that the SF does not saturate (e.g., J0502$+$1338). 
In such cases, we are unable to estimate the characteristic timescale $\tau_{\rm char}$, but can only place a lower limit on it, i.e., that the characteristic timescale is larger than the total observing time span.
For the 16 sources which do not show significant variability above $D_{\rm noise}$ (e.g., figure~\ref{J1350+0940}), the estimated SF is lower than the $D_{\rm noise}$ (figure~\ref{J1350+0940} lower panel), and $\tau_{\rm char}$ has a negative value derived by the SF model fit. For these sources, we consider $D_{\rm noise}$ as the upper limit of $D(\rm 1000d)$, and exclude them from all of the following variability amplitude and timescale analysis.

All the parameters of the OVRO sample used in the study, including variability amplitudes, characteristic timescale, redshift (from \citealp{2014MNRAS.438.3058R} and \citealp{2015MNRAS.452.4274P}), variability Doppler factors (from \citealp{2018ApJ...866..137L}), and multi-frequency core sizes (from \citealp{2022MNRAS.515.1736K}) are given in table~\ref{variability amplitude, timescale, and doppler factor} and table~\ref{core size data}.

\begin{table*}
        \begin{center}
            \caption{List of the variability amplitudes, characteristic timescales.
            Where no measurement is available, we place a dash in the respective column.
            (This table is available in its entirely in machine-readable form.)
            }
            \begin{tabular}{|c|c|c|c|c|c|c|c|c|c|c|} \hline
                Source name & $D(\rm 1000 d)$ & $D(\rm 1000 d)_{up}$ & $D(\rm 1000 d)_{low}$ & $\tau_{\rm char}$  & $\tau_{\rm{char, \ up}}$ & $\tau_{\rm{char, \ low}}$ & $D(\rm 4 d)$ & $D(\rm 4 d)_{up}$ & $D(\rm 4 d)_{low}$  \\ 
                (J2000) &  &  &  & [days] & [days] & [days] &  &  &   \\ 
                (1) & (2) & (3) & (4) & (5) & (6) & (7) & (8) & (9) & (10)  \\ \hline\hline
                J0001+1914 & 0.0791 & 0.0812 & 0.0771 & - & - & 3392 & 0.002774 & 0.002783 & 0.002764 \\
                J0001-1551 & 0.0752 & 0.0774 & 0.0730 & 1258 & 1437 & 1080 & 0.003532 & 0.00354 & 0.003523 \\
                J0003+2129 & 0.0177 & 0.0192 & 0.0162 & 1215 & 1762 & 668 & 0.022737 & 0.022816 & 0.022658 \\
                J0004+2019 & 0.0361 & 0.0379 & 0.0342 & 243 & 300 & 186 & 0.001982 & 0.001989 & 0.001976 \\
                J0004+4615 & 0.1546 & 0.1585 & 0.1508 & 299 & 336 & 262 & 0.004886 & 0.004901 & 0.004871 \\
                J0004-1148 & 0.0944 & 0.0977 & 0.0911 & 291 & 339 & 245 & 0.001333 & 0.001339 & 0.001328 \\
                J0005+0524 & 0.0060 & 0.0064 & 0.0056 & 3280 & 10877 & 3138 & 0.011481 & 0.011519 & 0.011444 \\
                J0005+3820 & 0.0434 & 0.0447 & 0.0422 & 147 & 169 & 126 & 0.002368 & 0.002378 & 0.002359 \\
                J0005-1648 & 0.0342 & 0.0352 & 0.0332 & - & - & 3032 & 0.006802 & 0.006824 & 0.006780 \\ 
                J0136+4751 & 0.1709 & 0.1769 & 0.1649 & 250 & 295 & 207 & 0.002604 & 0.002609 & 0.002599 \\ \hline
            \end{tabular}
            {\parbox{6.1 in}{
            Column notes: (2) Variability amplitude at 1000 days ($D(\rm 1000 d)$); (3) upper limit of $D(\rm 1000 d)$ at 95\% confidence level; (4) lower limit of $D(\rm 1000 d)$ at 95\% confidence level ; (5) characteristic timescale ($\tau_{\rm char}$); (6) upper limit of $\tau_{\rm char}$  at 95\% confidence level ; (7) lower limit of $\tau_{\rm char}$ at 95\% confidence level ; (8) variability amplitude at 4 days ($D(\rm 4 d)$); (9) 1$\sigma$ upper uncertainty of $D(\rm 4 d)$; (10) 1$\sigma$ lower uncertainty of $D(\rm 4 d)$.}
            }
            \label{variability amplitude, timescale, and doppler factor}
        \end{center}
    \end{table*}
    
    \begin{table*}
        \begin{center}
            \caption{
            List of the redshifts, variability Doppler factors and median of the observed (angular) core sizes from 1 to 86~GHz in the unit of milli-arcsecond.
            Where no measurement is available, we place a dash in the respective column.
            (This table is available in its entirely in machine-readable form.)
            }
            \begin{tabular}{|c|c|c|c|c|c|c|c|c|c|c|c|c|} \hline
                Source name & z & $\delta_{\rm{var}}$ & $\delta_{\rm{var, \ up}}$ & $\delta_{\rm{var, \ low}}$ & $\theta_{\rm{1\ GHz}}$ & $\theta_{\rm{2\ GHz}}$ & $\theta_{\rm{5\ GHz}}$ & $\theta_{\rm{8\ GHz}}$ & $\theta_{\rm{15 \ GHz}}$ & $\theta_{\rm{24 \ GHz}}$ & $\theta_{\rm{43 \ GHz}}$ & $\theta_{\rm{86 \ GHz}}$ \\
                (J2000) &  &  &  &  & [mas] & [mas] & [mas] & [mas] & [mas] & [mas] & [mas] & [mas]   \\ 
                (1) & (2) & (3) & (4) & (5) & (6) & (7) & (8) & (9) & (10) & (11) & (12) & (13)   \\ \hline\hline
                J0001+1914 & 3.1 & 1.82 & 30.76 & 0.50 & - & 0.6008 & 0.3010 & 0.1859 & - & - & - & - \\
                J0001-1551 & 2.044 & 2.51 & 1.69 & 1.44 & - & 0.6670 & - & 0.4600 & - & - & - & - \\
                J0003+2129 & 0.45 & - & - & - & - & - & - & 0.3220 & - & - & - & - \\
                J0004+2019 & 0.677 & 1.37 & 24.59 & 0.48 & - & - & - & 0.2975 & 0.2100 & - & - & - \\
                J0004+4615 & 1.81 & 7.75 & 1.70 & 2.75 & - & 0.7660 & - & 0.1743 & - & - & - & - \\
                J0004-1148 & 0.86 & - & - & - & - & 0.8760 & - & 0.4938 & - & - & - & - \\
                J0005+0524 & 1.887 & - & - & - & - & 1.2390 & - & 0.4235 & - & - & - & - \\
                J0005+3820 & 0.229 & 5.23 & 2.72 & 1.22 & - & 0.7618 & - & 0.2933 & 0.1356 & - & - & - \\
                J0005-1648 & 0.775 & - & - & - & - & 0.9020 & - & 0.4095 & - & - & - & - \\    
                J0136+4751 & 0.859 & 12.73 & 48.7 & 3.17 & 3.3980 & 0.9649 & 0.6187 & 0.1981 & 0.0992 & 0.0585 & 0.0630 & 0.0400 \\ \hline
            \end{tabular}
            {\parbox{6.5 in}{
            Note: (2) redshift \citep{2014MNRAS.438.3058R, 2015MNRAS.452.4274P}; (3) variability Doppler factor $\delta_{\rm{var}}$ \citep{2018ApJ...866..137L}; (4) upper limit of $\delta_{\rm{var}}$; (5) lower limit of $\delta_{\rm{var}}$; (6) 1~GHz core size; (7) 2~GHz core size; (8) 5~GHz core size; (9) 8~GHz core size; (10) 15~GHz core size; (11) 24~GHz core size; (12) 43~GHz core size; (13) 86~GHz core size. 1 to 86~GHz core sizes are taken from \citet{2022MNRAS.515.1736K}. }
            }
            \label{core size data}
        \end{center}
    \end{table*}

\subsection{Interpretation of Variability as Source-Intrinsic} 
\label{Interpretation of Variability as Source-Intrinsic}

Refractive interstellar scintillation \citep{1986ApJ...307..564R, 1999ApJS..121..483B, 1999ApJ...514..249B, 1999ApJ...514..272B} is known to cause variability at timescales from weeks to months at the centimeter wavelengths.
To confirm that the variability amplitudes at longer timelags in our following analyses are actually representative of the intrinsic variability of the sources and not due to refractive scintillation, we conducted correlation tests between $D(\rm 4d)$, $D(\rm 12d)$, $D(\rm 24d)$, $D(\rm 52d)$, $D(\rm 100d)$, $D(\rm 500d)$, and $D(\rm 1000d)$ (all multiples of the 4 day bins) with that of the line-of-sight H$-\rm \alpha$ intensity obtained from the Wisconsin H-Alpha Mapper (WHAM, \citealp{2003ApJS..149..405H}), which functions as a proxy for the amount of scattering material in the ionized ISM towards each source.
For $D(\rm 4d)$, $D(\rm 12d)$, $D(\rm 24d)$, and $D(\rm 52d)$, we used the measured values from equation~\ref{structure function}, while $D(\rm 100d)$, $D(\rm 500d)$, and $D(\rm 1000d)$ were derived from the SF model fitting from equation~\ref{SF 1000 days}.
We find that only $D(\rm 4d)$, $D(\rm 12d)$, and $D(\rm 24d)$ exhibit significant correlation with the the H$-\rm \alpha$ intensity ($r=0.12\pm0.02$ with $p=1.85\times10^{-5}$, $r=0.12\pm0.03$ with $p=5.73\times10^{-5}$, and $r=0.10\pm0.02$ with $p=4.06\times10^{-4}$).
However, variability amplitudes at timescale longer than 52~days, the correlation with H$-{\rm \alpha}$ is not significant. We can therefore conclude that the variability amplitudes at time lags longer than 52~days, including $D(\rm 1000d)$ which we use for our subsequent analyses, are dominated by intrinsic processes in the blazars themselves.

\subsection{Dependence of Variability Amplitudes on Core Sizes} 
\label{ssec: Dependence of Variability Amplitudes on Observed/Linear Core Sizes}

In this sub-section, we examine if the SF amplitudes at 1000 days are dependent on the milli-arcsecond core sizes measured at different frequencies using VLBI. 

\subsubsection{Dependence of Variability Amplitudes on Observed Angular Core Sizes} \label{Dependence of Variability Amplitudes on Observed Core Sizes}

\begin{figure*} 
    \centering
    \includegraphics[width=\textwidth]{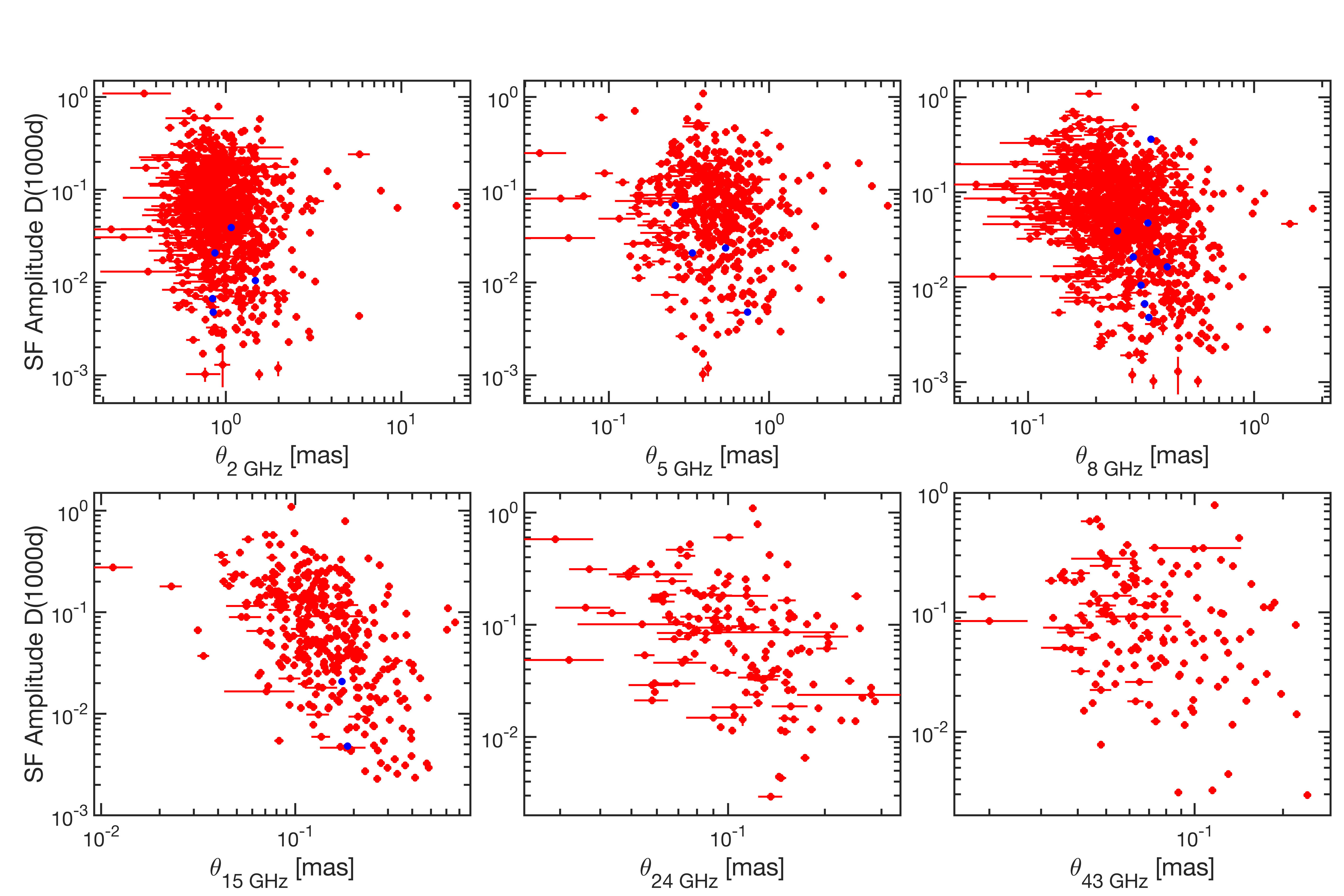}
    \caption{Scatter plot of SF amplitude at 1000 days ($D(\rm 1000d)$) against observed core size at milliarcsecond scales, measured at various frequencies.
    The error bars show the 1$\sigma$ uncertainties of SF amplitudes and angular core sizes.}
    Blue symbols are sources for which the fitted value of $D(\rm 1000d)$ have negative values, due to the variability being lower than $D_{\rm noise}$, such that the upper limit of $D(\rm 1000d)$ is given by $D_{\rm noise}$.
    The variability amplitudes show a dependence on core sizes, particularly that measured at higher frequencies.
    \label{D1000 vs Appsource}
\end{figure*}

Figure~\ref{D1000 vs Appsource} shows the scatter plot of the 15\,GHz $D(\rm 1000 d)$ against the observed angular core sizes, in units of milli-arcsecond, measured by \citet{2022MNRAS.515.1736K} at multiple frequencies. 
We see a weak anti-correlation between $D(\rm 1000 d)$ and the observed core sizes ($\theta$) measured at higher frequencies, particularly between 8 to 24\,GHz.
This relationship is less apparent by eye at low frequencies (i.e., 2 and 5\,GHz). 

We use the Spearman correlation test to quantify the strength of the relationship between two variables, in this case $D(\rm 1000 d)$ and the observed core sizes at various frequencies, and the significance of such a correlation. 
Table~\ref{table of correlation test} test A shows results of the correlation tests between $D(\rm 1000 d)$ and the observed core sizes at six different frequencies.

\begin{table}
    \centering
    \caption{
        Spearman correlation test results performed in this study. 
        Test A is the correlation test between $D(\rm 1000d)$ and observed core sizes~($\theta$) at 6 frequencies; test B is the correlation test between $D(\rm 1000d)$ and linear core sizes~($l$) at 6 frequencies; test C is the correlation test between $D(\rm 1000d)$ and intrinsic ($\theta_{ \rm{int, 1~GHz} }$)/scattering ($\theta_{ \rm{sca, 1~GHz} }$)/observed ($\theta_{ \rm{obs, 1~GHz} }$) core sizes at 1~GHz; test D is the correlation test between $D(\rm 4d)$ and intrinsic ($\theta_{ \rm{int, 1~GHz} }$)/scattering ($\theta_{ \rm{sca, 1~GHz} }$)/observed ($\theta_{ \rm{obs, 1~GHz} }$) core sizes at 1~GHz.
        Sample size, Spearman correlation coefficient, uncertainties of Spearman correlation coefficient, and $p$-value of each test are given in the table.
        Boldface $p$-values are considered significant.
        The uncertainties of correlation coefficient are derived from bootstrap resampling.
        }
    \begin{tabular}{|c|c|c|c|} \hline
            Test A & Sample & Correlation & $p$-value  \\  
                 & size & coefficient &   \\  \hline
            $D(\rm 1000d)$ v.s. $\theta_{\rm 2 \ GHz}$ & 860 & $-0.14 \pm 0.03$ & $\mathbf{1.82 \times 10^{-5}}$ \\
            $D(\rm 1000d)$ v.s. $\theta_{\rm 5 \ GHz}$ & 460 & $-0.13 \pm 0.04$ & $\mathbf{3.50 \times 10^{-3}}$ \\
            $D(\rm 1000d)$ v.s. $\theta_{\rm 8 \ GHz}$ & 1044 & $-0.35 \pm 0.03$ & $\mathbf{3.28 \times 10^{-33}}$ \\
            $D(\rm 1000d)$ v.s. $\theta_{\rm 15 \ GHz}$ & 375 & $-0.52 \pm 0.05$ & $\mathbf{1.74 \times 10^{-27}}$ \\
            $D(\rm 1000d)$ v.s. $\theta_{\rm 24 \ GHz}$ & 156 & $-0.43 \pm 0.08$ & $\mathbf{1.74 \times 10^{-8}}$ \\
            $D(\rm 1000d)$ v.s. $\theta_{\rm 43 \ GHz}$ & 158 & $-0.23 \pm 0.07$ & $\mathbf{3.20 \times 10^{-3}}$ \\ \hline\hline
            Test B & Sample  & Correlation & $p$-value  \\  
                  &  size & coefficient &   \\ \hline
            $D(\rm 1000d)$ v.s. $l_{\rm 2 \ GHz}$ & 764 & $-0.10 \pm 0.03$  & $\mathbf{3.10 \times 10^{-3}}$ \\
            $D(\rm 1000d)$ v.s. $l_{\rm 5 \ GHz}$ & 422 & $-0.14 \pm 0.04$  & $2.61 \times 10^{-2}$ \\
            $D(\rm 1000d)$ v.s. $l_{\rm 8 \ GHz}$ & 912 & $-0.32 \pm 0.02$ & $ \mathbf{\approx 0}$ \\
            $D(\rm 1000d)$ v.s. $l_{\rm 15 \ GHz}$ & 344 & $-0.39 \pm 0.04$ & $\mathbf{1.52 \times 10^{-14}}$ \\
            $D(\rm 1000d)$ v.s. $l_{\rm 24 \ GHz}$ & 147 & $-0.32 \pm 0.08$ & $\mathbf{6.13 \times 10^{-5}}$ \\
            $D(\rm 1000d)$ v.s. $l_{\rm 43 \ GHz}$ & 150 & $-0.14 \pm 0.08$ & $7.30 \times 10^{-2}$ \\ 
            \hline\hline
            Test C & Sample & Correlation & $p$-value  \\  
                  &  size & coefficient &   \\ \hline
            $D(\rm 1000d)$ v.s. $\theta_{ \rm{int, 1~GHz} }$ & 536 & $-0.29 \pm 0.04$ & $\mathbf{5.34 \times 10^{-12}}$ \\
            $D(\rm 1000d)$ v.s. $\theta_{ \rm{sca, 1~GHz} }$ & 230 & $-0.06 \pm 0.06$ & $3.18 \times 10^{-1}$ \\
            $D(\rm 1000d)$ v.s. $\theta_{ \rm{obs, 1~GHz} }$ & 536 & $-0.06 \pm 0.04$ & $1.21 \times 10^{-1}$ \\ \hline\hline
            Test D & Sample & Correlation & $p$-value  \\  
                  & size & coefficient &   \\ \hline
            $D(\rm 4d)$ v.s. $\theta_{ \rm{int, 1~GHz} }$ & 538 & $-0.21 \pm 0.03$ & $\mathbf{8.48 \times 10^{-7}}$ \\
            $D(\rm 4d)$ v.s. $\theta_{ \rm{sca, 1~GHz} }$ & 230 & $-0.07 \pm 0.06$ & $2.69 \times 10^{-1}$ \\
            $D(\rm 4d)$ v.s. $\theta_{ \rm{obs, 1~GHz} }$ & 538 & $-0.09 \pm 0.04$ & $\mathbf{3.49 \times 10^{-2}}$ \\ \hline
    \end{tabular}
    \label{table of correlation test}
\end{table}

\begin{table*}
    \centering
    \begin{flushleft}
        \caption{
        Results of two sample K-S test. 
        Test A is the K-S test comparing the distribution of observed core sizes of weak and strong variables at 6 frequencies; test B is the K-S test comparing the distribution of linear core sizes of weak and strong variables at 6 frequencies; test C is the K-S test comparing the distribution of observed core sizes of slow and fast variables at 6 frequencies; test D is the K-S test comparing the distribution of linear core sizes of slow and fast variables at 6 frequencies; test E is the K-S test comparing the distribution of intrinsic/scattering/observed core sizes of weak and strong variables ($D(\rm 1000 d)$) at 1~GHz; test F is the K-S test of distribution of intrinsic/scattering/observed core sizes of slow and fast variables; test G is the K-S test of distribution of intrinsic/scattering/observed core sizes of weak and strong variables ($D(\rm 4 d)$) at 1~GHz.
        Sample size, median value of each test, median core sizes of each distribution, the unit of the core sizes are shown in the brackets, and $p$-value of each test are given in the table. 
        Boldface $p$-values are considered significant.
        }
    \begin{tabular}{|c|c|c|c|c|c|} \hline
            Test A & Sample & Median of  & Median $\theta$ of & Median $\theta$ of & $p$-value \\
            (Separated by median of $D(\rm 1000d)$) & size & $D(\rm 1000d)$ & weak variables [mas] & strong variables [mas]  \\  \hline
            Distribution of $\theta_{\rm 2 \ GHz}$ of weak and strong variables & 864 &  0.0604 & 0.99 & 0.91 & $\mathbf{1.83 \times 10^{-2}}$ \\
            Distribution of $\theta_{\rm 5 \ GHz}$ of weak and strong variables & 464 & 0.0605 & 0.49 & 0.45 & $1.30 \times 10^{-1}$ \\
            Distribution of $\theta_{\rm 8 \ GHz}$ of weak and strong variables & 1053 & 0.0601 & 0.31 & 0.24 & $\mathbf{7.77 \times 10^{-17}}$ \\
            Distribution of $\theta_{\rm 15 \ GHz}$ of weak and strong variables & 376 & 0.0691 & 0.18 & 0.12 & $\mathbf{2.58 \times 10^{-15}}$ \\
            Distribution of $\theta_{\rm 24 \ GHz}$ of weak and strong variables & 156 & 0.0801 & 0.13 & 0.09 & $\mathbf{1.58 \times 10^{-6}}$ \\
            Distribution of $\theta_{\rm 43 \ GHz}$ of weak and strong variables & 158 & 0.0920 & 0.07 & 0.06 & $5.78 \times 10^{-2}$ \\ \hline\hline
            
            Test B & Sample & Median of & Median $l$ of & Median $l$ of & $p$-value \\
            (Separated by median of $D(\rm 1000d)$) & size & $D(\rm 1000d)$ & weak variables [pc] & strong variables [pc] \\  \hline
            Distribution of $l_{\rm 2 \ GHz}$ of weak and strong variables & 768 &  0.0604 & 7.24 & 6.61 & $\mathbf{3.52 \times 10^{-2}}$ \\
            Distribution of $l_{\rm 5 \ GHz}$ of weak and strong variables & 424 &  0.0605 & 3.72 & 3.01 & $\mathbf{1.80 \times 10^{-3}}$ \\
            Distribution of $l_{\rm 8 \ GHz}$ of weak and strong variables & 917 &  0.0601 & 2.24 & 1.74 & $\mathbf{7.30 \times 10^{-15}}$ \\
            Distribution of $l_{\rm 15 \ GHz}$ of weak and strong variables & 346 & 0.0691 & 1.22 & 0.79 & $\mathbf{6.15 \times 10^{-14}}$ \\
            Distribution of $l_{\rm 24 \ GHz}$ of weak and strong variables & 147 & 0.0801 & 1.00 & 0.63 & $\mathbf{3.56 \times 10^{-5}}$ \\
            Distribution of $l_{\rm 43 \ GHz}$ of weak and strong variables & 150  & 0.0920 & 0.52 & 0.42 & $7.18 \times 10^{-2}$ \\ \hline\hline
            
            Test C & Sample & Median of  & Median $\theta$ of & Median $\theta$ of & $p$-value \\
            (Separated by median of $\tau_{\rm char}$) & size & $\tau_{\rm char}$ [days] & slow variables [mas] & fast variables [mas] \\  \hline
            Distribution of $\theta_{\rm 2 \ GHz}$ of slow and fast variables & 864 & 650 & 0.99 & 0.91 & $\mathbf{4.72 \times 10^{-4}}$ \\
            Distribution of $\theta_{\rm 5 \ GHz}$ of slow and fast variables & 463 & 637 & 0.50 & 0.44 & $1.78 \times 10^{-1}$ \\
            Distribution of $\theta_{\rm 8 \ GHz}$ of slow and fast variables & 1053 & 630 & 0.25 & 0.28 & $\mathbf{1.07 \times 10^{-4}}$ \\
            Distribution of $\theta_{\rm 15 \ GHz}$ of slow and fast variables & 376 & 560 & 0.16 & 0.12 & $\mathbf{2.33 \times 10^{-6}}$ \\
            Distribution of $\theta_{\rm 24 \ GHz}$ of slow and fast variables & 155 & 544 & 0.12 & 0.10 & $\mathbf{4.52 \times 10^{-2}}$ \\
            Distribution of $\theta_{\rm 43 \ GHz}$ of slow and fast variables & 158 & 482 & 0.07 & 0.06 & $6.33 \times 10^{-2}$ \\ \hline\hline
            
            Test D & Sample & Median of  & Median $l$ of & Median $l$ of & $p$-value \\
            (Separated by median of $\tau_{\rm src}$) & size & $\tau_{\rm src}$ [days] & slow variables [pc] & fast variables [pc] \\  \hline
            Distribution of $l_{\rm 2 \ GHz}$ of slow and fast variables & 590 & 2515 & 7.15 & 6.30 & $\mathbf{2.54 \times 10^{-3}}$ \\
            Distribution of $l_{\rm 5 \ GHz}$ of slow and fast variables & 348 & 2815 & 3.65 & 3.08 & $\mathbf{1.38 \times 10^{-2}}$ \\
            Distribution of $l_{\rm 8 \ GHz}$ of slow and fast variables & 674 & 2454 & 1.92 & 1.73 & $\mathbf{4.35 \times 10^{-3}}$ \\
            Distribution of $l_{\rm 15 \ GHz}$ of slow and fast variables & 312 & 2786 & 0.98 & 0.84 & $\mathbf{2.30 \times 10^{-2}}$ \\
            Distribution of $l_{\rm 24 \ GHz}$ of slow and fast variables & 144 & 3160 & 0.84 & 0.77 & $7.40 \times 10^{-1}$ \\
            Distribution of $l_{\rm 43 \ GHz}$ of slow and fast variables & 152 & 3181 & 0.44 & 0.46 & $1.35 \times 10^{-1}$ \\ \hline\hline
            
            Test E & Sample & Median of & Median core size of & Median  core size of & $p$-value \\
            (Separated by median of $D(\rm 1000d)$) & size & $D(\rm 1000d)$ & weak variables [mas] & strong variables [mas] \\  \hline
            Distribution of $\theta_{ \rm{int, 1~GHz} }$ of weak and strong variables & 536 & 0.0601 & 2.04 & 1.66 & $\mathbf{3.58 \times 10^{-6}}$ \\
            Distribution of $\theta_{ \rm{sca, 1~GHz} }$ of weak and strong variables & 230 & 0.0601 & 3.55 & 3.47 & $9.35 \times 10^{-1}$ \\
            Distribution of $\theta_{ \rm{obs, 1~GHz} }$ of weak and strong variables & 536 & 0.0601 & 2.75 & 2.63 & $3.16 \times 10^{-1}$ \\ \hline\hline

            Test F & Sample & Median of  & Median core size of & Median core size of & $p$-value \\
            (Separated by median of $\tau_{\rm char}$) & size & $\tau_{\rm char}$ [days] & slow variables [mas] & fast variables [mas] \\  \hline
            Distribution of $\theta_{ \rm{int, 1~GHz} }$ of slow and fast variables & 536 & 633 d & 1.99 & 1.74 & $\mathbf{2.45 \times 10^{-3}}$ \\
            Distribution of $\theta_{ \rm{sca, 1~GHz} }$ of slow and fast variables & 230 & 550 d & 3.71 & 3.50 & $1.48 \times 10^{-1}$ \\
            Distribution of $\theta_{ \rm{obs, 1~GHz} }$ of slow and fast variables & 536 & 633 d & 2.72 & 2.70 & $4.99 \times 10^{-1}$ \\ \hline\hline
            
            Test G & Sample & Median of & Median core size of & Median core size of & $p$-value \\
            (Separated by median of $D(\rm 4 d)$) & size & $D(\rm 4 d)$ & weak variables [mas] & strong variables [mas] \\  \hline
            Distribution of $\theta_{ \rm{int, 1~GHz} }$ of weak and strong variables & 538 & 0.0015 & 1.95 & 1.70 & $\mathbf{1.27 \times 10^{-3}}$ \\
            Distribution of $\theta_{ \rm{sca, 1~GHz} }$ of weak and strong variables & 230 & 0.0015 & 3.63 & 3.46 & $4.31 \times 10^{-1}$ \\
            Distribution of $\theta_{ \rm{obs, 1~GHz} }$ of weak and strong variables & 538 & 0.0015 & 2.63 & 2.88 & $8.17 \times 10^{-2}$ \\ \hline\hline
    \end{tabular}
    \label{two sample K-S test}
    \end{flushleft}
\end{table*}

We assume a significance level of $\alpha = 0.05$, such that $p$-values below 0.05 are defined as statistically significant results.
We find a significant anti-correlation, between the $D(1000 \rm d)$ and the observed core sizes measure at all frequencies. 

From table~\ref{table of correlation test} test A, we can see that the absolute correlation coefficients are smaller at lower frequencies, which indicates that the correlations are weak, but still significant. 
However, at 8, 15, and 24\,GHz, the absolute correlation coefficients are higher, and $p$-values are much lower.
Since $D(1000 \rm d)$ values are derived from the lightcurves at 15\,GHz, we expect the strongest correlation with the observed core sizes at this frequency, which is exactly what we observe, where the correlation coefficient is the highest. 
However, we obtain a much smaller $p$-value at 8\,GHz due to the larger sample with observed core size measurements at that frequency.

\begin{figure*} 
    \centering
    \includegraphics[width=\textwidth]{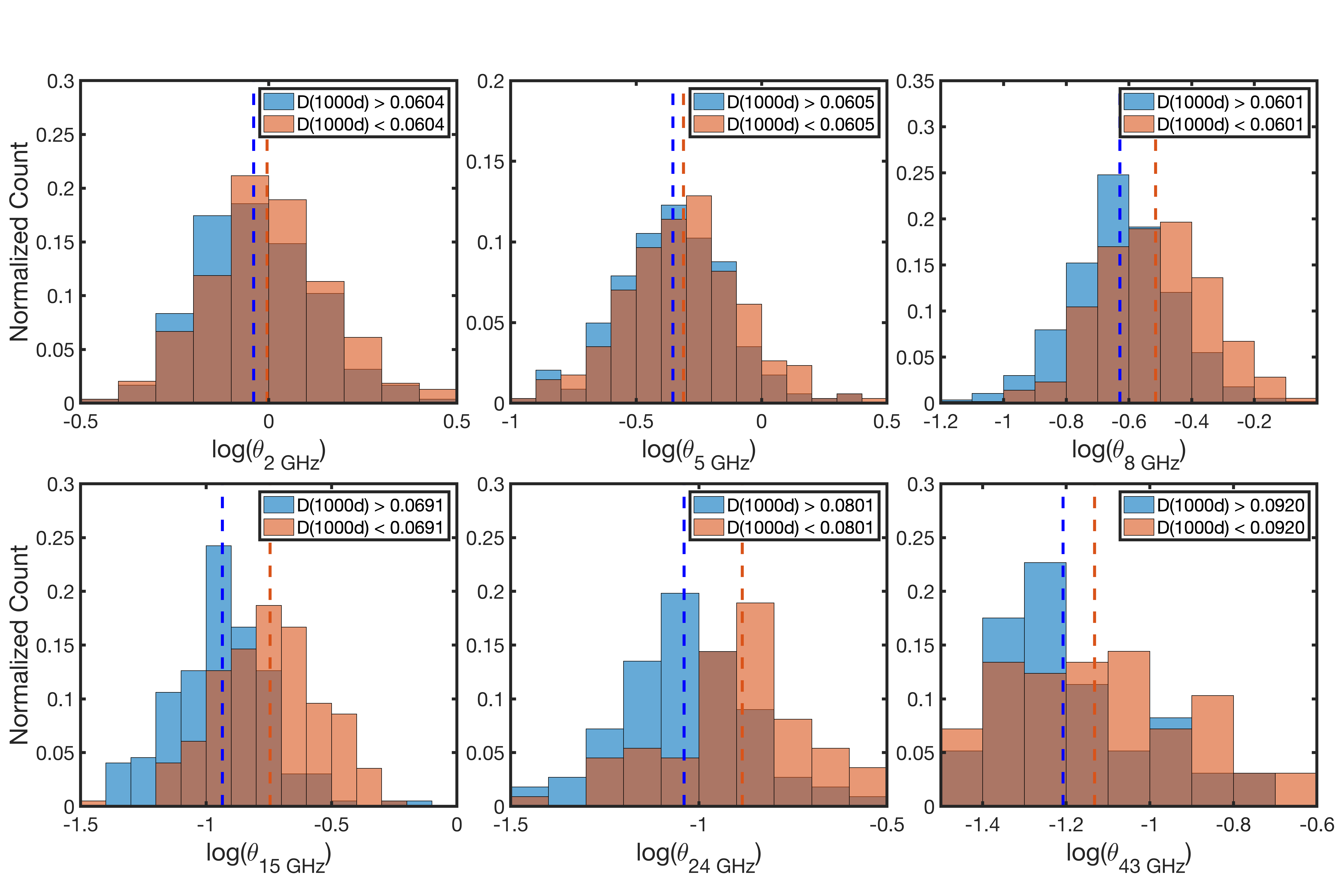}
    \caption{
    Distributions of the observed core sizes of the strong and weak variables, separated by the median of $D(\rm 1000d)$ for each frequency sample.
    The blue and orange dashed lines show the median value of the observed core sizes of the strong and weak variables, respectively.
    We find that the core sizes of strong variables are significantly smaller than that of the weak variables, particularly for core sizes measured at 8\,GHz and higher.
    The brown areas correspond to overlapping regions between the blue and orange histograms.
    }
    \label{D1000 vs Observed source distribution}
\end{figure*}

Additionally we also use the two-sample Kolmogorov-Smirnov (K-S) test to compare the distributions of the observed core sizes of both strong and weak variables to see if they are significantly different. 
At each frequency for which we have observed core size measurements, the number of overlapping sources from the OVRO sample is different. 
Therefore, for each frequency, we separate the overlapping OVRO sources into two sub-samples, by the median of $D(\rm 1000d)$. 
Note that the median of $D(\rm 1000d)$ is different for samples with observed core sizes measured at different frequencies, as shown in table~\ref{two sample K-S test} test A. 
At each frequency, we refer to all sources with $D(\rm 1000d)$ larger than the median value as the strong variables, while those with $D(\rm 1000d)$ lower than the median value are referred to as the weak variables. 
After separating the samples into strong and weak variables for each of the frequencies with observed core size measurements, we compare the core size distributions of the two samples.

Figure~\ref{D1000 vs Observed source distribution} shows the distributions of the observed core sizes measured at the different frequencies, for both the strong and weak variables. 
We see that the strong variables (blue histograms) have typically smaller observed core sizes; while the weak variables (orange histograms) have larger observed core sizes.
The median values (see table~\ref{two sample K-S test} test A for more details) of the observed core sizes of weak and strong variables are more comparable at low frequencies (i.e., 2 and 5\, GHz), e.g., $\tilde{\theta}_{2 \ \rm GHz, weak}$ is about 9\% larger than $\tilde{\theta}_{2 \ \rm GHz, strong}$; $\tilde{\theta}_{15 \ \rm GHz, weak}$ is approximately 33\% larger than $\tilde{\theta}_{15 \ \rm GHz, strong}$.

With the exception of the 43\,GHz observed core sizes, the K-S tests show a significant difference between the observed core size distributions of the strong and weak variables. Additionally, even though at 2\,GHz the $p$-value is low enough to be classified as significant, it is marginal. 

\subsubsection{Dependence of Variability Amplitudes on Linear Core Sizes} 
\label{Dependence of Variability Amplitudes on Linear Core Sizes}

We expect that the intrinsic variability characteristics should be dependent on the physical size rather than angular size, which we examine next.
The linear core sizes ($l$) were calculated from the observed (angular) core sizes using the equation, 
\begin{equation}
\label{linear core sizes}
   l = \theta \times d_{\rm{A}}
\end{equation}
where $l$ is the linear/physical core size, in units of parsec, $\theta$ is the observed core size, and $d_{\rm{A}}$ is the angular diameter distance.
We assume the standard cosmological constants as follows: Hubble constant $H_0 = 68 \ \rm km \ s^{-1} \ Mpc^{-1}$, $\Omega_{\rm m} = 0.3$, and $\Omega_ {\rm \lambda} = 0.7$.
We calculate the linear core sizes for all sources with redshift information. 
For sources with no redshifts, we excluded them from the following analysis.

\begin{figure*} 
    \centering
    \includegraphics[width=\textwidth]{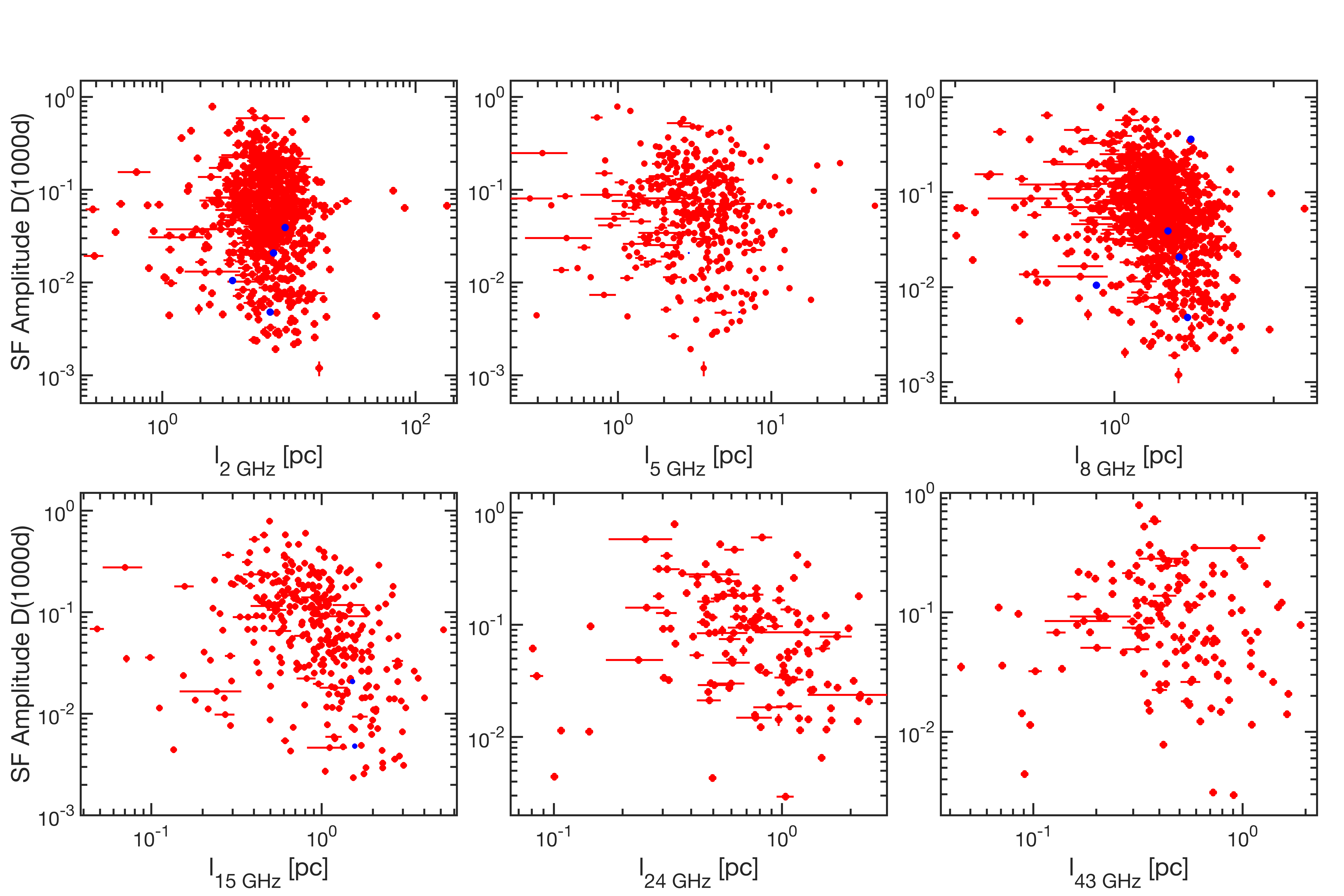}
    \caption{Scatter plot of SF amplitude at 1000 days ($D(\rm 1000d)$) against linear core size in units of parsecs. 
    The error bars show the 1$\sigma$ uncertainties of SF amplitudes and linear core sizes.}
    Blue symbols are sources for which the fitted value of $D(\rm 1000d)$ have negative values, due to the variability being lower than $D_{\rm noise}$, such that the upper limit of $D(\rm 1000d)$ is given by $D_{\rm noise}$.
    The physical core sizes generally increase as the SF amplitude at 1000 days decreases. 
    \label{D1000 vs linear core sizes}
\end{figure*}

Figure~\ref{D1000 vs linear core sizes} shows the scatter plot of $D(\rm 1000d)$ against the linear (physical) core sizes ($l$), for sources for which redshift measurements are available. 
As can be seen in table~\ref{table of correlation test} test B, the results are consistent with those found when the observed angular core sizes were used. 
The anti-correlation between $D(\rm 1000d)$ and the linear core sizes is still statistically significant. 
The $p$-values are slightly larger at low frequencies, likely due to the smaller sample sizes for sources with redshift information. 

\begin{figure*} 
    \centering
    \includegraphics[width=\textwidth]{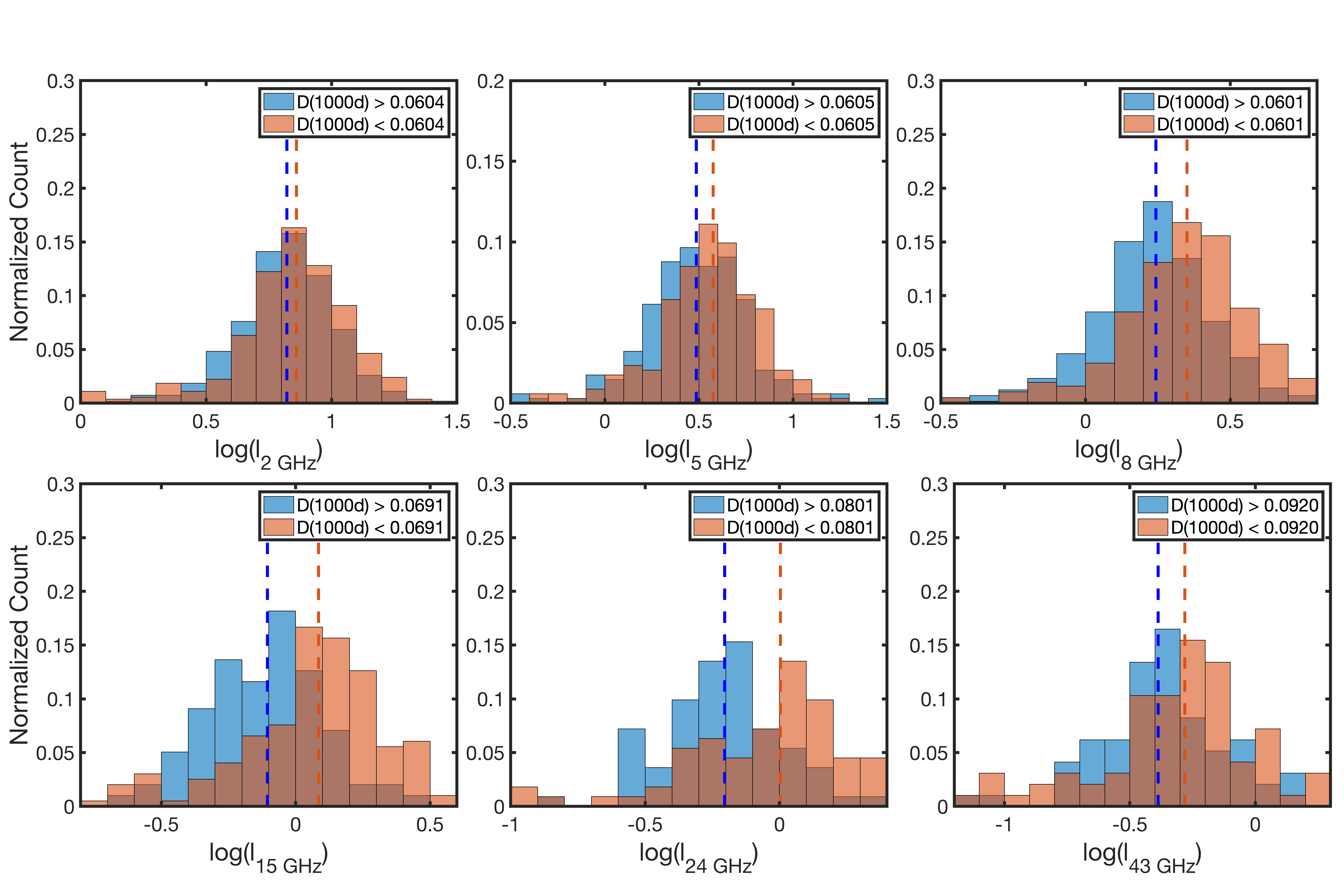}
    \caption{
    The distributions of physical core size at different frequencies, separated into strong and weak variables by the median of the variability amplitudes $D(\rm 1000d)$.
    The blue and orange dashed lines show the median value of the physical core sizes of the strong and weak variables, respectively.
    The brown areas correspond to overlapping regions between the blue and orange histograms.
    }
    \label{D1000 vs linear source distribution}
\end{figure*}

Similarly, we perform further analysis by comparing the physical core size distributions for the strong and weak variables. 
Figure~\ref{D1000 vs linear source distribution} shows the distributions of the physical core sizes measured at different frequencies, separating the sample into strong and weak variables by the median of $D(\rm 1000 d)$. 
The strong variable sources (blue histograms) have more compact linear core sizes than the weak variables (orange histograms).
Also, the median linear core sizes of strong variables (blue dashed lines) are smaller than the weak variable sources (orange dashed lines) at all frequencies.

In table~\ref{two sample K-S test} test B, all $p$-values except at 43\,GHz are lower than 0.05, which shows that the linear core sizes of the strong variables are significantly smaller than that of the weak variables.
The median of the linear core sizes are close at low frequencies (i.e., 2 and 5\,GHz), e.g., $\tilde{l}_{2 \ \rm GHz, weak}$ is about 9\% larger than $\tilde{l}_{2 \ \rm GHz, strong}$; $\tilde{\theta}_{15 \ \rm GHz, weak}$ is approximately 35\% larger than $\tilde{\theta}_{24 \ \rm GHz, strong}$.

\subsubsection{Discussion of the Variability Amplitude and Core Size Relationship} 
\label{Discussion of Dependence of Variability Amplitudes on Core Sizes}

From sections~\ref{Dependence of Variability Amplitudes on Observed Core Sizes} and \ref{Dependence of Variability Amplitudes on Linear Core Sizes}, we found significant correlation of $D(\rm 1000 d)$ with the angular and linear core sizes between 8 to 24~GHz.
Stronger variables have significantly smaller core sizes relative to weaker variables.
Although the light curves were measured at 15~GHz, significant results were also found between the variability amplitudes and core sizes observed at other frequencies including 8 and 24 GHz. 
This can be explained by a combination of opacity effects, and if the jet structure is frequency dependent, such as typically assumed in the conical jet model \citep{1979ApJ...232...34B}, an idealized model of a steady radio jet of AGN. In this model, there is a relationship between core size and observed frequency due to opacity effects, given by 
 \begin{equation}
    \label{conical jet model}
   \theta_{\rm size} \propto \nu^{-1} ,
\end{equation}
where $\theta_{\rm size}$ is the observed core size, and $\nu$ is the observed frequency. 
At higher frequency observations, the $\tau = 1$ surface shifts upstream, and one probes a narrower region of the jet compared to that at lower frequencies. 
The fact that we observe significant correlations between the 15\,GHz variability amplitudes and the core sizes measured at other frequencies (including 2 and 8\,GHz), points to this possible relationship between core size and observing frequency.

However, there are likely to be deviations from an idealized conical steady-state jet model, due to the clumpy and complex structure of jets. 
As we measure core sizes at increasingly lower or higher frequencies, we expect the deviations from an idealised jet to increase.
This can explain why, for observed and linear core sizes observed at lower or higher frequencies, i.e., 2, 5 and 43~GHz, the significance level ($p$-value) and strength of correlation decreases. 

At low frequencies of 2 and 5\,GHz, the decrease in significance of the results can also be explained by interstellar scattering affecting the observed core size measurements. 
From their full sample of sources, \citet{2022MNRAS.515.1736K} found a significant correlation between the 2~GHz observed core sizes with H$-\rm \alpha$ intensity, which traces the column density of ionized gas in the Galaxy. 
To confirm that this effect is also present in our sub-sample of sources overlapping with the OVRO sample, we examine if the observed core sizes at 2\,GHz for our sample also show significant correlations with the line-of-sight H$-\rm \alpha$ intensities.
Using the Kendall $\rm \tau$ correlation test, following the paper of \citet{2022MNRAS.515.1736K}, we found that there is significant correlation of the 2~GHz observed core sizes (both in low galactic latitude (|b|$\le$15) sources, and full sample) with the WHAM H$-\rm \alpha$ intensity.
The correlation coefficient and $p$-value of low galactic latitude (N = 71) and the full sample (N = 579) is $\rm \tau = 0.29 \pm 0.10$, $p = 4.25 \times 10^{-4}$, and $\rm \tau = 0.12 \pm 0.04$, $p = 3.03 \times 10^{-5}$, respectively.

Note that $D(\rm 1000d)$ is determined in the observer's frame, such that for higher redshift sources, the variability amplitudes are measured at much shorter time lags in the rest frame of the source. 
This introduces biases to the analysis, since it is more likely that the higher redshift sources have SFs that have yet to saturate.
To determine if this affects our conclusions, we perform the same analysis using the $D(\rm 1000d)$ in the rest frame of the source by fitting the following SF model, 
\begin{equation}
    \label{rest frame SF 1000 days}
    D_{\rm mod}(\tau) = D\left(\frac{\rm 1000d}{1+z}\right) \frac{1+(1+z)\tau_{\rm char}/1000 }{1+\tau_{\rm char}/\tau}+ D_{\rm noise}
\end{equation}
where $D\left(\frac{\rm 1000d}{1+z}\right)$ is the variability amplitude at 1000 days at the source rest frame, $D_{\rm rest}(\rm 1000d)$.
We estimated the $D_{\rm rest}(\rm 1000d)$ of 982 OVRO sources with redshift measurements. 
We found a significant relationship between $D_{\rm rest}(\rm 1000d)$ and the linear core sizes at all observed frequencies, consistent with our results using the observed frame $D(\rm 1000d)$. This confirms that the relationship between the variability amplitudes and linear core sizes is robust, regardless of whether we use $D_{\rm rest}(\rm 1000d)$ or $D(\rm 1000d)$.  

\subsection{Dependence of Variability Timescale on Core Sizes} 
\label{Dependence of Variability Timescale on Observed/Linear Core Sizes}

In this sub-section, we examine the relationship between the characteristic and intrinsic variability timescales (corresponding to timescales in the observer's frame and the source rest frame respectively) of the blazars in our sample and their core sizes (both angular and linear).

\subsubsection{Dependence of Characteristic Timescale on Angular Core Sizes} 
\label{Dependence of Variability Timescale on Observed Core Sizes}

Due to the fact that we can only obtain a lower limit on the variability timescale of a large fraction (306/1157) of sources (due to the SF not saturating within the observing timespan), we do not perform correlation tests in this section. 
For each frequency with observed core size measurements, we separate the source sample into fast and slow variables, based on the median of the characteristic timescales, $\tau_{\rm char}$. 
The former is defined as sources with $\tau_{\rm char}$ shorter than the median characteristic timescale in the sample, while the latter is defined as sources with characteristic timescales longer than the median value of the sample.

\begin{figure*} 
    \centering
    \includegraphics[width=\textwidth]{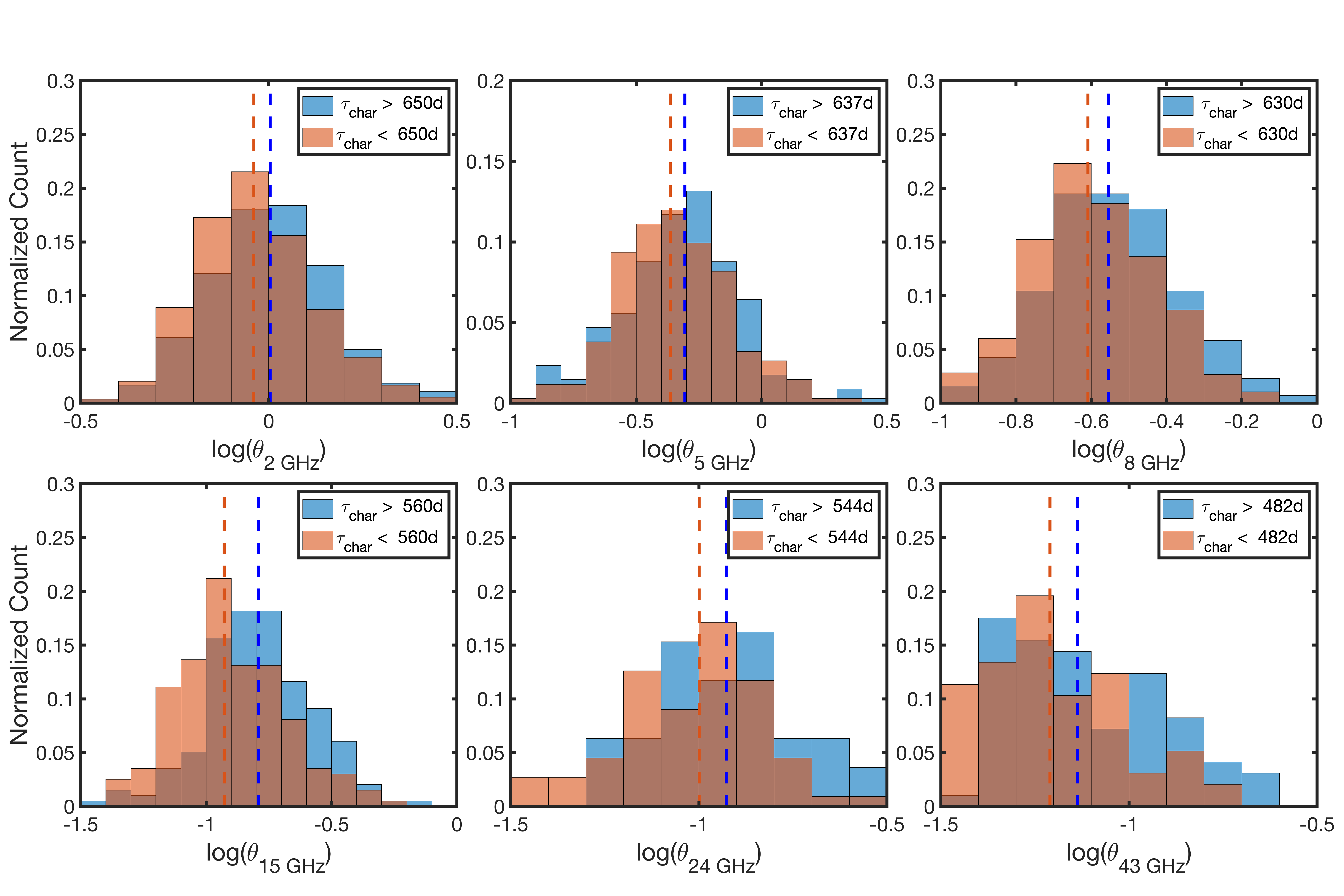}
    \caption{
    Distributions of observed core size of slow and fast variables, which is separated by the median of characteristic timescales $\tau_{\rm char}$. 
    The blue and orange dashed lines show the median of core sizes of the slow and fast variables at each frequency where the core size is measured.
    The brown areas correspond to overlapping regions between the blue and orange histograms.
    }
    \label{characteristic timescale vs observed source distribution}
\end{figure*}

Figure~\ref{characteristic timescale vs observed source distribution} compares the distributions of observed core angular sizes of sources with short characteristic timescales in the frame of the observer (fast variables, orange histograms) and that with long characteristic variability timescales (slow variables, blue histograms).

For core sizes measured at all frequencies, we see that the fast variables typically have smaller observed core sizes, while slow variables have larger observed core sizes. 
The median values of $\tau_{\rm char}$ separating the fast and slow variables for each frequency sample can be seen in table~\ref{two sample K-S test} test C. 
Also, the corresponding two sample K-S test results are shown in table~\ref{two sample K-S test} test C, where we determine if the distribution of the observed core sizes of the fast and slow variables are drawn from the same population.
For all frequencies except for 5 and 43\,GHz, we see that the observed core sizes of the slow variables are significantly larger than that of the fast variables. 

\subsubsection{Dependence of Intrinsic Timescale on Linear Core Sizes} 
\label{Dependence of Variability Timescale on Linear Core Sizes}

We expect the intrinsic variability timescales at the rest frame of the source to show a stronger relationship with the intrinsic linear core sizes. 
Additionally, we need to consider time dilation effects for sources at cosmological distances, as well as time compression effects due to Doppler beaming. 
To calculate the intrinsic timescale (variability timescale in the rest frame of the source), we use the equation given by:
\begin{equation}
    \label{intrinsic timescale}
    \tau_{\rm src} = \frac{\delta}{(1+z)}\tau_{\rm char} 
\end{equation}
where $\tau_{\rm src}$ is the variability timescale at the rest frame of the source, $\delta$ is Doppler boosting factor, taken from \citet{2018ApJ...866..137L} for sources for which data are available, $\tau_{\rm char}$ is the characteristic timescale that we determined from the SF fitting, which we derived in section~\ref{sssec: Structure Function Fitting}, and $z$ is the source redshift. 

\begin{figure*} 
    \centering
    \includegraphics[width=\textwidth]{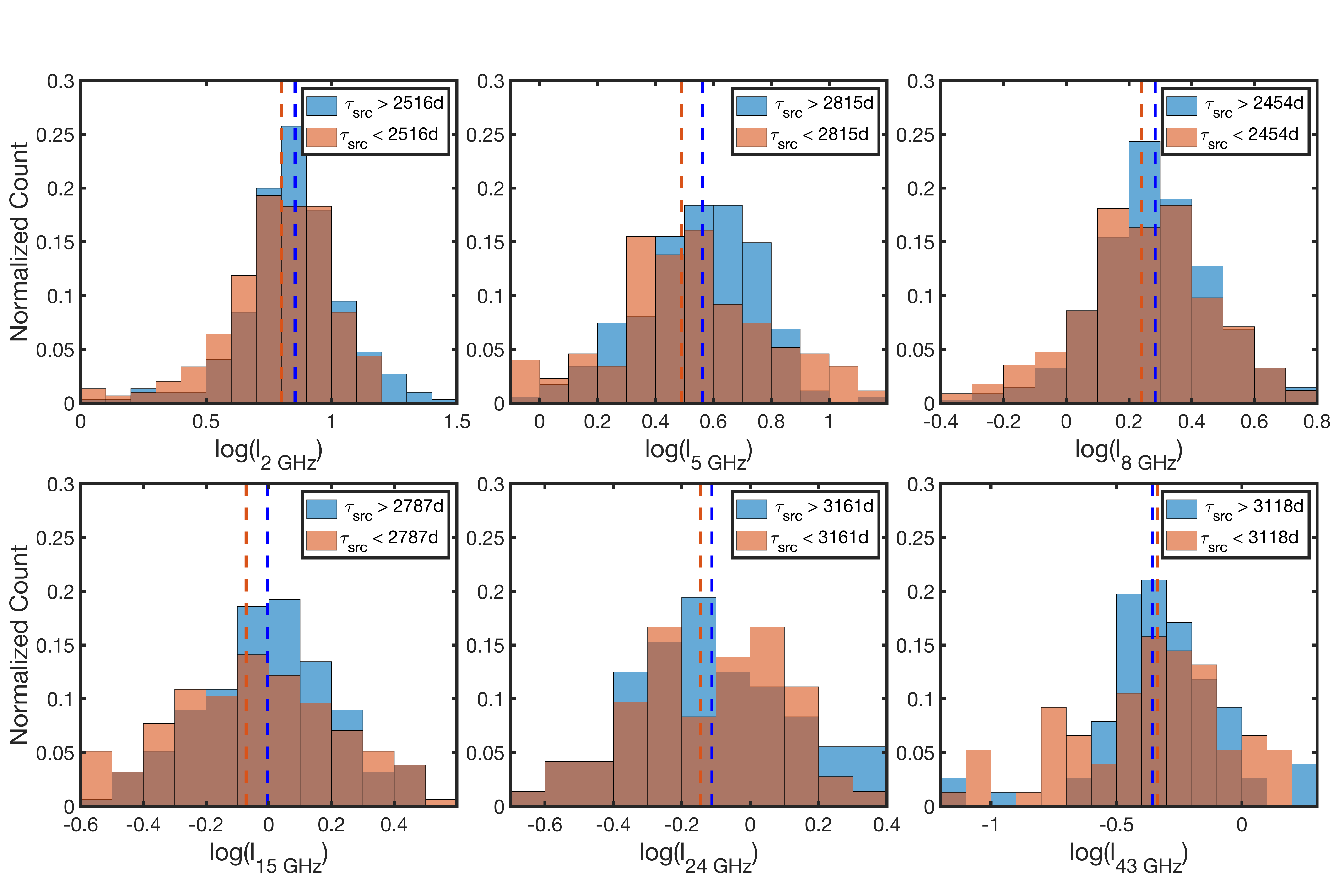}
    \caption{
    The distribution of linear core sizes measured at various frequencies, separated by the median of the intrinsic timescale $\tau_{\rm src}$.
    The blue and orange dashed lines are the median value of physical core sizes of the slow and fast variables, respectively.
    The brown areas correspond to overlapping regions between the blue and orange histograms.
    }
    \label{intrinsic timescale vs linear source distribution}
\end{figure*}

We now examine the dependence of these derived intrinsic timescales on the linear core sizes derived in section~\ref{Dependence of Variability Amplitudes on Linear Core Sizes}. 
At each frequency with core size measurements, we separate the sources into two samples, by the median value of the intrinsic timescale ($\tau_{\rm src}$).
Figure~\ref{intrinsic timescale vs linear source distribution} shows the distributions of linear core sizes for sources with long intrinsic timescales (slow variables) and short intrinsic timescales (fast variables). 
Distributions from figure~\ref{intrinsic timescale vs linear source distribution} are similar to that in figure~\ref{characteristic timescale vs observed source distribution}, where sources with fast intrinsic variability timescales tend to have smaller linear core sizes, while sources with slow intrinsic variation timescale tend to have larger physical core sizes. 
Also, the median values of the linear core sizes are similar at low frequencies, e.g., $\tilde{l}_{2 \ \rm GHz, slow}$ is about 12\% larger than $\tilde{l}_{2 \ \rm GHz, fast}$; $\tilde{l}_{15 \ \rm GHz, slow}$ is approximately 14\% larger than $\tilde{l}_{15 \ \rm GHz, fast}$.
The median values of the linear core sizes at each frequency, for the fast and slow variables, can be see at table~\ref{two sample K-S test} test D. 

The corresponding two sample K-S test results are shown in table~\ref{two sample K-S test} test D.
We find that the slow variables have significantly larger linear core sizes than the fast variables, for core sizes measured at 2 to 15\,GHz.
The $p$-values have generally increased for correlations between the intrinsic timescale and the 2, 8, and 15\,GHz linear core sizes compared to the $p$-values from table~\ref{two sample K-S test} test C, likely due to the smaller number for sources with both redshift and Doppler boosting factor information.
On the other hand, the distributions of the 5\,GHz linear core sizes of the slow and fast variables are now significantly different, compared to that in test C when the characteristic timescales in the observer's frame and the angular core sizes are used.

\subsubsection{Discussion on the Dependence of Variability Timescale on Core Sizes} 
\label{Discussion of Dependence of Variability Timescale on Core Sizes}

As seen with the correlation of variability amplitudes and observed core sizes in section~\ref{ssec: Dependence of Variability Amplitudes on Observed/Linear Core Sizes}, the two sample K-S test results are most significant around 8 and 15~GHz (table~\ref{two sample K-S test} test C and D), because the light curves were observed at 15~GHz. 
The level of significance decreases for core sizes measured at lower and higher frequencies. 
As explained in section~\ref{Discussion of Dependence of Variability Amplitudes on Core Sizes}, this can be explained by a combination of opacity effects and the conical jet model such that core sizes have a frequency dependence.

This dependence of variability timescale on linear sources sizes can be explained based on the light travel time argument. 
For 15~GHz source classified as slow variables, the median intrinsic timescale is 7129 days ($\approx$ 19.53~years), and the median 15\,GHz linear core size is approximately 0.98~pc ($\approx$3.19~ly).
The median core size of slow variable sources is smaller than the upper limit of the variability timescale.
On the other hand, for the fast variable sources, the median intrinsic timescale is about 1169~days ($\approx$ 3.2~years), and the median 15\,GHz linear core size is 0.84~pc ($\approx$ 2.73~ly). 
For the fast variable sources, the median linear core size is consistent with the light travel-time argument, in fact they appear on average to vary at the limit imposed by the speed of light.

\subsection{Examining the Effects of Interstellar Scattering at 1\,GHz} 
\label{Dependence of Variability Amplitudes and Timescales on Intrinsic/Scattering/Observed Core Sizes}

\subsubsection{Derivation of Intrinsic and Scattering Sizes from the Observed Core Sizes} 
\label{Derivation of Intrinsic/Scattering/Observed Core Sizes}

In this sub-section, we argue that the correlation between variability amplitudes and core sizes measured at lower frequencies is weaker due to the measured angular core sizes being affected by interstellar scattering.     
The core sizes of compact AGN will be broadened by interstellar scattering at low frequencies \citep{2008ApJ...672..115L,2015MNRAS.452.4274P, 2022MNRAS.515.1736K} where the effects of scattering are most significant.

We examine again the dependence of variability amplitudes and timescales on core sizes measured at low frequencies, specifically at 1~GHz. 
This time, however, we look at the intrinsic core sizes after the scattering component has been separated. 
This separation of the intrinsic core size and scatter broadening component was performed by \citet{2022MNRAS.515.1736K}, where they studied the frequency dependence of the core sizes to study synchrotron opacity in AGN jets and the strength of scatter broadening in the ISM at different lines of sight in our Milky Way. 
The frequency dependence of the observed core size and observed frequency relationship can be describe as 
\begin{equation}
    \label{core size-frequency dependence}
   \theta_{\rm size} \propto \nu^{-k}
\end{equation}
where $\theta_{\rm size}$ is the observed core size, $\nu$ is the observed frequency, and the $k$ index describes the power law relationship between core size and observed frequency. They derive $k$ for each source based on fitting the above equation to the multi-frequency core size measurements from VLBI observations. 
For sources above the Galactic plane ($|b|>10^{\circ}$), they found that the observed angular core sizes show a frequency dependence of $ \theta_{\rm size} \propto \nu^{-1.02 \pm 0.01}$, which is consistent with the conical jet model of \citet{1979ApJ...232...34B}. 
For sources observed through the Galactic plane between $\pm 10^{\circ}$, they found that the distributions of the $k$ values show two separate peaks, one at $-0.99 \pm 0.02$, and another at $-1.60 \pm 0.02$. 
This result suggests that there are two populations of sources. 
One population has core sizes showing a frequency dependence of $ \theta_{\rm size} \propto \nu^{-0.99 \pm 0.02}$ consistent with that expected for the conical jet model and thus not affected significantly by interstellar scattering. 
The other population shows a steeper relationship of $\theta_{\rm size} \propto \nu^{-1.60 \pm 0.02}$, where scattering effects from the ISM along the line of sight are significant.

However, the theoretically expected $k$ value at low Galactic latitudes should be about $2$ \citep{1986ApJ...310..737C, 1990ARA&A..28..561R, 1995ApJ...443..209A}, assuming a Kolmogorov power spectrum.
However, the observed core size is actually the convolution of the intrinsic core size and the size of the scattering disk (angular broadening component), which can be estimated by 
\begin{equation}
    \label{observed core size equation}
    \theta^2_{\rm obs} = \theta^2_{\rm int} + \theta^2_{\rm sca}
\end{equation}
where $\theta_{\rm obs}$ is observed core size, $\theta_{\rm int}$ is intrinsic core size, and $\theta_{\rm sca}$ is the size of the scattering disk.
\citet{2022MNRAS.515.1736K} fit two power law functions to the plot of the core size as a function of frequency, each with a different k-index, using equation~\ref{core size-frequency dependence}. One function represents the scattering size, and another represents the intrinsic core size at each frequency.  
With these fits, they derive both the intrinsic core size and the size of the scattering disk at 1\,GHz. 

We use these intrinsic and scattering sizes in table~7 of \citet{2022MNRAS.515.1736K} for our following analysis.
There are 538 sources for which they have derived scattered and intrinsic core sizes that are also overlapping with our OVRO sample. 
Out of these, 497 sources have redshift information.
In the following two sub-sections, we discuss the dependence of the 1\,GHz intrinsic, scattered and observed core sizes with the long term intrinsic variability $D(\rm 1000d)$ and the short term variability $D(\rm 4d)$.

\subsubsection{Dependence of Long Term Intrinsic Variability on the Intrinsic, Scattering, and Observed Core Sizes} 
\label{Dependence of Variability Amplitudes at 1000 days on Intrinsic/Scattering/Observed Core Sizes} 

Here we use the SF amplitude at 1000 days ($D(\rm 1000 d)$) derived using equation~\ref{SF 1000 days} as described in section~\ref{sssec: Structure Function Fitting}.

\begin{figure*} 
    \centering
    \includegraphics[width=\textwidth]{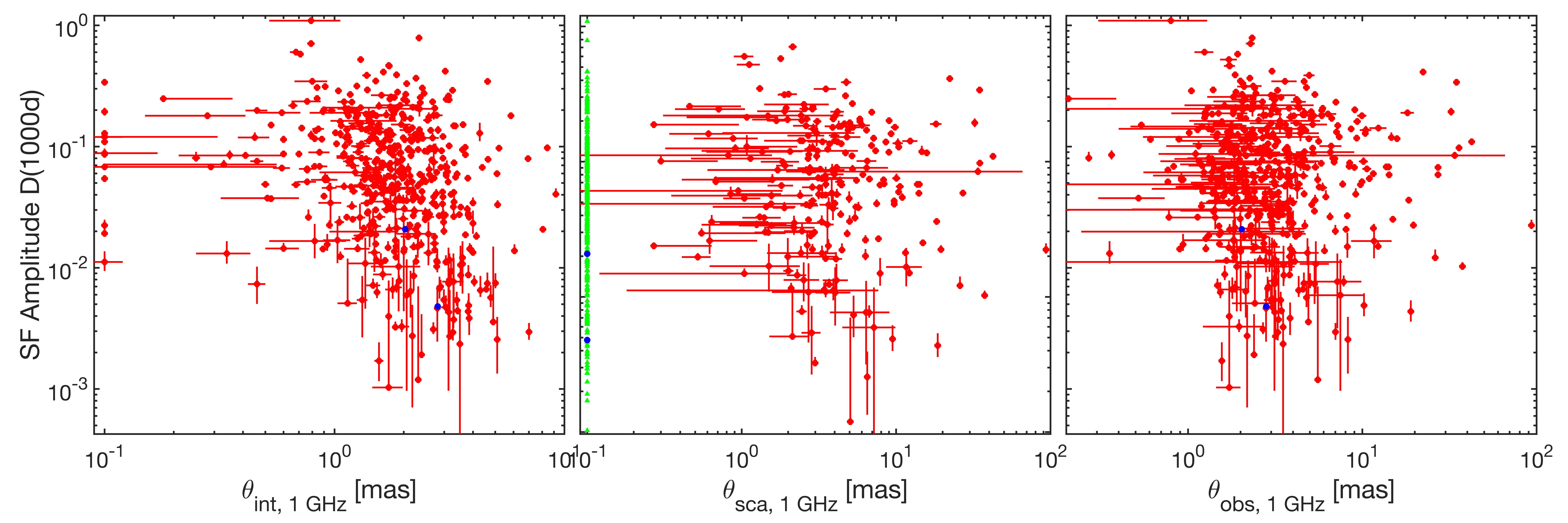}
    \caption{Scatter plot of SF amplitude at 1000 days $D(\rm 1000d)$ against 1~GHz intrinsic core sizes (left panel), scattering core sizes (middle panel), and observed core sizes (right panel), in units of milli-arcseconds.
    The error bars shows the uncertainty of SF amplitudes, and core size uncertainties.
    Blue symbols are sources for which the fitted value of $D(\rm 1000d)$ have negative values, due to the variability being lower than $D_{\rm noise}$, such that the upper limit of $D(\rm 1000d)$ is given by $D_{\rm noise}$.
    Green triangles in the middle panel represent upper limits of the scattering sizes, indicating that the effects of scattering are insignificant for these sources, which are not included in the correlation test.
    Observed core sizes are derived from equation~\ref{observed core size equation}.
    }
    \label{D1000 vs intrinsic/scattering/observed core size}
\end{figure*}

Figure~\ref{D1000 vs intrinsic/scattering/observed core size} shows the scatter plot of $D(\rm 1000 d)$ against the 1\,GHz intrinsic core sizes (left panel), scattering sizes (middle panel), and observed core sizes (right panel).
The corresponding correlation test results can be seen in table~\ref{table of correlation test} test~C.
Sources for which the scattering sizes are insignificant (green triangles in the middle panel of figure~\ref{D1000 vs intrinsic/scattering/observed core size}) are excluded from the correlation test.
We only see a significant correlation between $D(\rm 1000 d)$ and the intrinsic core sizes (left panel), with a correlation coefficient of about $-0.29 \pm 0.04$, and $p$-value of $5.34 \times 10^{-12}$.
$D(\rm 1000 d)$ shows no significant correlation with the scattering core size and the observed core size. 

\begin{figure*} 
    \centering
    \includegraphics[width=\textwidth]{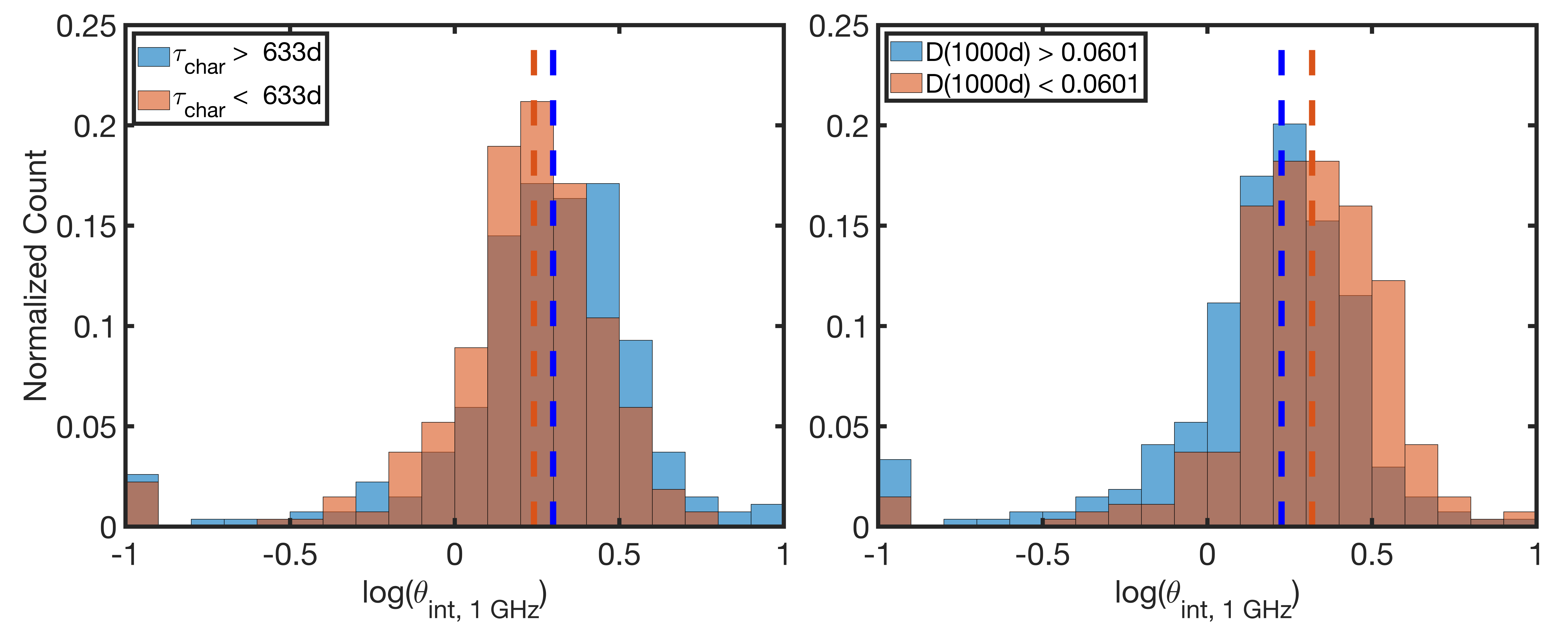}
    \caption{Comparing the distributions of 1~GHz intrinsic core sizes ($\theta_{\rm{int, 1~GHz}}$) of the fast (orange histogram) and slow (blue histogram) variables, separated by the median of $\tau_{\rm char}$ (left panel), and that of the weak (orange histogram) and strong (blue histogram) variables, separated by the median of $D(\rm 1000 d)$ (right panel).
    Dashed lines are the median values of the logarithm of $\theta_{\rm{int, 1~GHz}}$ of each sample. The brown areas correspond to overlapping regions between the blue and orange histograms.
    }
    \label{intrinsic core size separation}
\end{figure*}

Next we separate the intrinsic, scattering, and observed core sizes at 1 GHz into two samples, based on whether they correspond to weak or strong variables, separated by the median of the variability amplitudes $D(\rm 1000 d)$. 
The right panel of Figure~\ref{intrinsic core size separation} shows the  distribution of 1~GHz intrinsic core sizes of these weak and strong variables.
The two sample K-S test results are shown in table~\ref{two sample K-S test} test E, where we see that the intrinsic core sizes of weak and strong variables are significantly different.
We exclude those sources which have insignificant scattering effect, i.e., $\theta_{\rm{sca, 1~GHz}}=0.1 \ \rm{mas}$, remaining 230 sources. 
The strong variables (orange histograms) tend to have smaller $\theta_{\rm{int, 1~GHz}}$ than the weak variables (blue histograms).

These analyses demonstrate that at 1\,GHz when scattering is dominant, any relationship between variability amplitudes and measured core sizes will be weak or nonexistent, since sources observed through highly scattered sight-lines will be significantly broadened, and thus contaminating the relationship.
However, once we deconvoluted the scattering effects from the intrinsic core sizes, the correlation between $D(\rm 1000d)$ and intrinsic core size becomes significant, even at 1~GHz frequencies.

\subsubsection{Dependence of Variability Timescales on the Intrinsic, Scattering, and Observed Core Sizes} 
\label{Dependence of Variability Timescales on Intrinsic/Scattering/Observed Core Sizes} 

We again use the two sample K-S test to check if there is any significant difference between the 1\,GHz core sizes of slow and fast variables, by separating them based on the median of the characteristic timescale $\tau_{\rm char}$. 
The left panel of Figure~\ref{intrinsic core size separation} displays the distribution of the 1~GHz intrinsic core sizes of the fast and slow variables (separated by the median of $\tau_{\rm char}$).
The fast variable sources (orange histograms) tend to have smaller $\theta_{\rm{int, 1~GHz}}$. In contrast, the slow variable sources (blue histograms) have larger $\theta_{\rm{int, 1~GHz}}$.

Again for the test involving the scattering sizes, we exclude sources for which the scattering is insignificant. 
The results can be seen in table~\ref{two sample K-S test} test F. 
We found a significant difference between the distribution of the intrinsic core sizes of the slow and fast variables; no significant differences were found in the distributions of the scattering and observed core sizes of the weak and fast variables. 
Again, this supports the above argument that scattering effects weaken any relationship between core sizes measured at low frequencies and the variability characteristics, in this case the timescale of variability. 

\subsubsection{Dependence of Variability Amplitudes at 4 days on the Intrinsic, Scattering, and Observed Core Sizes}
\label{Dependence of Variability Amplitudes at 4 days on Intrinsic/Scattering/Observed Core Sizes}

The results in section~\ref{ssec: Dependence of Variability Amplitudes on Observed/Linear Core Sizes} and \ref{Dependence of Variability Timescale on Observed/Linear Core Sizes} apply only if the source of variability is dominated by processes intrinsic to the blazar itself. 
When the flux density variations are due to ISS, the results may be different, and we explore this in this section. 
As shown in section~\ref{Interpretation of Variability as Source-Intrinsic} and \citet{2019MNRAS.489.5365K}, $D(4 \rm d)$ is dominated by ISS.


We used equation~\ref{structure function} to calculate the SF amplitude of the 15~GHz OVRO lightcurves at a timelag of 4 days ($D(\rm 4 d)$), instead of using the SF model fits (equation~\ref{SF 1000 days}), which are more appropriate for characterising the long-term variations.

\begin{figure*} 
    \centering
    \includegraphics[width=\textwidth]{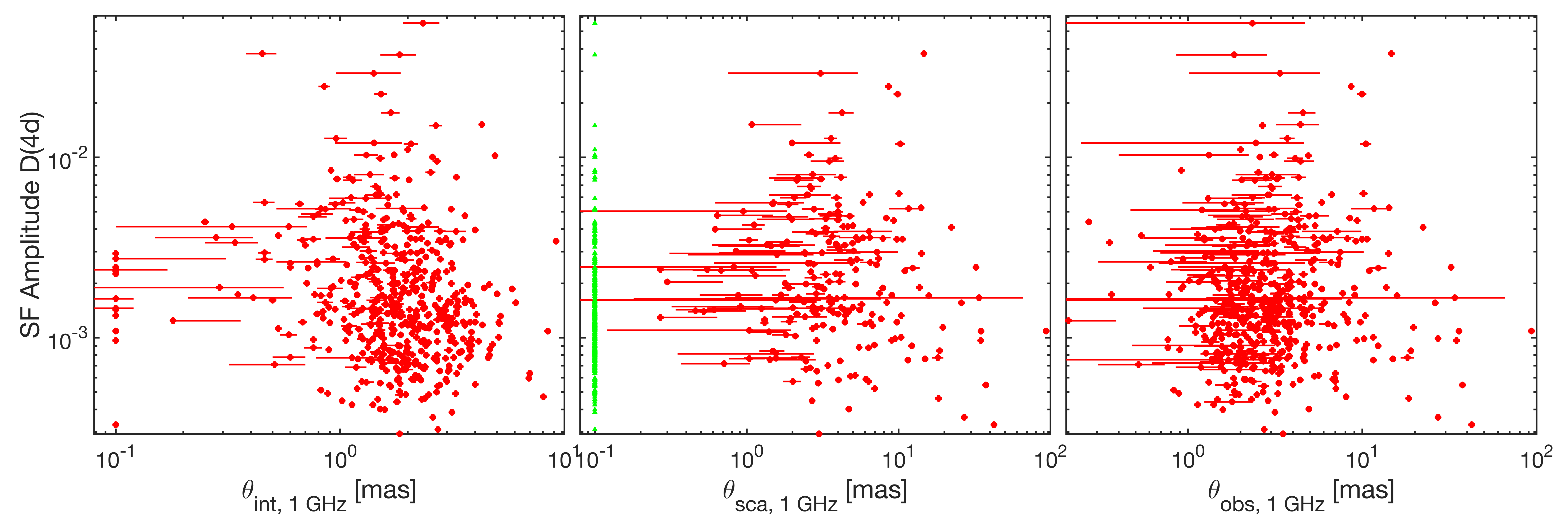}
    \caption{Scatter plot of SF amplitude at 4 days ($D(\rm 4 d)$) against 1~GHz intrinsic core sizes (left panel), scattering sizes (middle panel), and observed core sizes (right panel), in units of milli-arcsecond.
    The error bar shows the uncertainty of SF amplitudes, and core size uncertainties.
    Green triangles in the middle panel represent upper limits of the scattering sizes, which are not included in the correlation test.
    Observed core sizes are derive from equation~\ref{observed core size equation}.
    }
    \label{D4 vs intrinsic/scattering/observed core size}
\end{figure*}

Figure~\ref{D4 vs intrinsic/scattering/observed core size} displays the scatter plot of 15~GHz SF amplitude at 4 days $D(\rm 4 d)$ against the 1~GHz intrinsic core sizes (left panel), scattering core sizes (middle panel), and observed core sizes (right panel).
The corresponding correlation test results are displayed in table~\ref{table of correlation test} test~D. 
Upper limits of the scattering sizes (green triangles in the middle panel of figure~\ref{D4 vs intrinsic/scattering/observed core size}) are also excluded from the correlation test.

As with $D(\rm 1000 d)$, we find that only the 1~GHz intrinsic core sizes are significantly anti-correlated with $D(\rm 4 d)$, with a correlation coefficient of $-0.21 \pm 0.04$, and $p = 8.48 \times 10^{-7}$, indicating that sources with more compact structure exhibit higher amplitude $D(\rm 4 d)$ than those with more extended intrinsic core sizes.
Though the dependence is slightly weaker than that at $D(\rm 1000 d)$ (to be expected since scintillation amplitudes are also dependent on the line-of-sight scattering properties), the result is still significant. We see the same consistent result using the K-S tests and separating the sources into strong and weak variables (see table~\ref{two sample K-S test} test G), and the results of our analysis are consistent with the study by \citet{2005AJ....130.2473K}. 

The fact that the scintillation amplitudes are correlated only with the intrinsic core size and not the observed sizes is interesting. We know that the strength of ISS is strongly dependent on the source angular size, where more compact sources are known to scintillate more strongly than extended sources. 
This is the same reason why stars twinkle but planets do not in the optical regime due to atmospheric scintillation. 
This result indicates that the scintillation is sensitive to the intrinsic core sizes, not the total angular size inclusive of the angular broadening from scattering. 
This in turn suggests that the screen responsible for scatter broadening is either located nearer to the observer compared to the screen responsible for the ISS, or is the same screen responsible for the ISS itself.

If the latter is true, and the scattering screen responsible for the 1\,GHz angular broadening is the same screen responsible for the ISS, one would expect to see a correlation between $D(\rm 4 d)$ and the scattering disk sizes, which is not seen in the correlation tests. 
We argue that this is because sources that have insignificant scattering have been excluded from the analysis, thereby biasing the results.

Following this, we investigate the difference in distribution of $D(\rm 4 d)$ between the sample of sources with significant scatter broadening ($\theta_{\rm sca,  1~GHz}>0.1~\rm mas$) and those with insignificant scattering ($\theta_{\rm sca, 1~GHz} < 0.1~\rm mas$). 
\begin{figure} 
    \centering
    \includegraphics[width=\columnwidth]{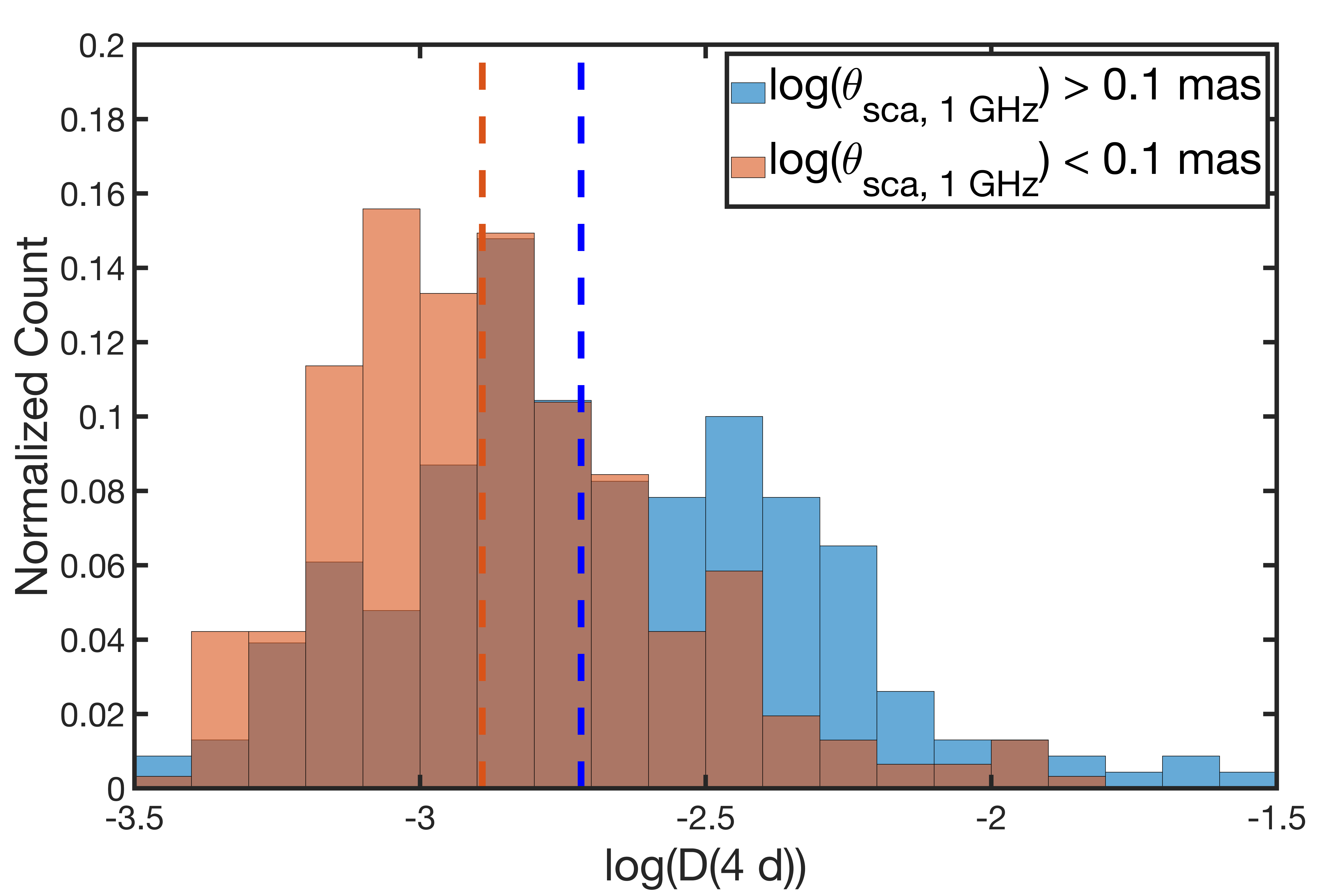}
    \caption{Distribution of logarithmic value of SF amplitude at 4 days ($D(\rm 4 d)$), separated by sources with significant scattering effect (blue histogram) and insignificant scattering effect (orange histogram).
    Dashed lines are the median value of the logarithmic value of $D(\rm 4 d)$ of each distribution.
    The brown areas correspond to overlapping regions between the blue and orange histograms.
    Core sizes lower than 0.1 mas are regarded as upper limit.
    }
    \label{D4 distribution}
\end{figure}
Figure~\ref{D4 distribution} shows the distribution of $D(\rm 4 d)$ of the two source samples, one with significant scattering and another with insignificant scatter broadening.
We again use the two sample K-S test to test the distribution difference.
Sources with insignificant scattering have significantly ($p=1.96 \times 10^{-8}$) lower $D(\rm 4 d)$ than those with significant scattering effects, indicating that sources that exhibit the strongest scintillation are also observed through sight lines where scatter broadening is also the most significant. 

\section{Conclusion}
\label{conclusion}

In this paper, we studied the 15~GHz variability of 1157 radio-selected blazars from the OVRO monitoring program, and their dependence on milliarcsecond core sizes measured at multiple frequencies.
Using the SF, we characterize the source variability amplitudes at a thousand days ($D(\rm 1000 d)$), the characteristic variability timescales ($\tau_{\rm char}$) of all sources, and the intrinsic variability timescales ($\tau_{\rm src}$) for sources where redshift information is available.
We summarize the main results of our study below:
\begin{enumerate}
    \item We found significant negative correlation between 15~GHz $D(\rm 1000 d)$ and the observed/linear core sizes at 2-24~GHz.
    The strong variables (with large variability amplitudes) have significantly smaller angular and linear core sizes compared to the weak variables. 
    This dependence is seen even using core sizes measured at other frequencies, likely due to the combination of opacity effects and the presence of a core size-frequency relation as expected from e.g., a conical jet.
    
    \item The fast variables with shorter characteristic and intrinsic variability timescales have significantly smaller angular (and linear) core sizes at 2-43~GHz, compared to slow variables sources. 
    This is consistent with what we expect based on the light travel time argument.
    
    \item We also confirmed that the correlation of 15~GHz $D(\rm 1000 d)$ and observed/linear core sizes at low frequencies is weaker or smeared out due to contamination of the observed/linear core sizes from ISM scattering. 
    After deconvolving scattering sizes, we found that the 15~GHz variability amplitudes and timescale show significant dependence on the 1~GHz intrinsic core sizes.
    
    \item Significant correlation was detected between 1~GHz intrinsic core sizes and short term variability amplitudes, i.e., 15~GHz $D(\rm 4 d)$. 
    Due to the fact that low-frequency core sizes are highly broadened by ISM, ISS strength is therefore dependent on the source intrinsic angular size. 
    Also, we found that strong variable sources have larger scattering core sizes.
\end{enumerate}

Blazars are known to exhibit structural changes on very short timescales from months to years \citep{2019MNRAS.485.1822P, 2023A&A...672A.130C}, such that their core sizes may vary, and may thus affect any such correlation studies if only a single epoch core size measurement is obtained for each source. 
We note, however, that our study uses the median core sizes measured over multiple years and decades \citep{2022MNRAS.515.1736K}, hence represent the `typical' core sizes of the sources in our sample.
Nevertheless, it will be interesting to conduct follow-up studies to examine changes in the variability characteristics of single sources as a function of changes in their core sizes or compactness over time. The OVRO lightcurves, combined with VLBI monitoring of the blazar structure, e.g. from the MOJAVE program \citep{2005AJ....130.1389L}, will be useful for such work.

\section*{Acknowledgements}
This research has made use of data from the OVRO 40-m monitoring program (Richards, J. L. et al. 2011, ApJS, 194, 29), supported by private funding from the California Institute of Technology and the Max Planck Institute for Radio Astronomy, and by NASA grants NNX08AW31G, NNX11A043G, and NNX14AQ89G and NSF grants AST-0808050 and AST-1109911.
P. C. H. is supported by the Taiwan National Science and Technology Council (NSTC, Grant No. NSTC 111-2124-M-001-005 and NSTC 110-2124-M-001-007).
T. H. was supported by the Academy of Finland projects 317383, 320085, 322535, and 345899.
S.K. acknowledges support from the European Research Council (ERC) under the European Unions Horizon 2020 research and innovation programme under grant agreement No.~771282.
W.M. gratefully acknowledges support by the ANID BASAL project FB210003 and FONDECYT 11190853.
We are also grateful to the anonymous referee for the helpful comments that helped improve the manuscript. 

\section*{Data availability}
\label{Data availability}

The derived data products including the structure function amplitudes and variability timescales are available in the article and in its online supplementary material. 
The light curve data are available upon request to the OVRO team, and are governed by the OVRO 40-m Data Usage Policy. 
See: \url{https://sites.astro.caltech.edu/ovroblazars/data.php?page=data_query} for more details.

\bibliographystyle{mnras}
\bibliography{reference}

\bsp
\label{lastpage}
\end{document}